\documentclass{nature}

\usepackage{graphicx}
\usepackage{amssymb,amsmath,bm}
 
\usepackage{float}

\usepackage{multirow}
\usepackage{threeparttable}
\usepackage{tablefootnote}
\usepackage{amsfonts, mathrsfs,xcolor}
\usepackage{lineno}
\usepackage{hyperref}
\usepackage{multirow}
\usepackage{booktabs}
\usepackage{geometry}
\usepackage{makecell} 
\usepackage{footnote}
\usepackage{float}
\usepackage{threeparttable}
\usepackage{xcolor}
\usepackage{tabularx}
\usepackage{amsmath}

\def\nat{Nature\ }
\def\aap{Astron.\ Astrophys.\ }
\def\apj{Astrophys.\ J.\ }
\def\apjl{Astrophys.\ J.\ Lett.\ }
\def\apjs{Astrophys.\ J.\ Supp.\ }

\def\mnras{Mon.\ Not.\ R.\ Astron.\ Soc.\ }

\def\prd{Phys.\ Rev.\ D\ }
\def\prl{Phys.\ Rev.\ Lett.\ }

\def\araa{Annu.\ Rev.\ Astron.\ Astrophys.\ }

\def\ssr{Space\ Sci.\ Rev.\ }

\begin{document}
\begin{Large}
\noindent
\textbf{An extreme particle accelerator powered by pulsar PSR~J1849-0001}
\end{Large}

\noindent
\begin{center}
Zhen Cao$^{1,2,3}$,
F. Aharonian$^{3,4,5,6}$,
Y.X. Bai$^{1,3}$,
Y.W. Bao$^{7}$,
D. Bastieri$^{8}$,
X.J. Bi$^{1,2,3}$,
Y.J. Bi$^{1,3}$,
W. Bian$^{7}$,
A.V. Bukevich$^{9}$,
C.M. Cai$^{10}$
% ...
(The LHAASO Collaboration)

$^{1}$ Key Laboratory of Particle Astrophysics \& Experimental Physics Division \& Computing Center, Institute of High Energy Physics, Chinese Academy of Sciences, 100049 Beijing, China\\
$^{2}$ University of Chinese Academy of Sciences, 100049 Beijing, China\\
$^{3}$ TIANFU Cosmic Ray Research Center, Chengdu, Sichuan,  China\\
$^{4}$ University of Science and Technology of China, 230026 Hefei, Anhui, China\\
$^{5}$ Yerevan State University, 1 Alek Manukyan Street, Yerevan 0025, Armeni a\\
$^{6}$ Max-Planck-Institut for Nuclear Physics, P.O. Box 103980, 69029  Heidelberg, Germany\\
$^{7}$ Tsung-Dao Lee Institute \& School of Physics and Astronomy, Shanghai Jiao Tong University, 200240 Shanghai, China\\
$^{8}$ Center for Astrophysics, Guangzhou University, 510006 Guangzhou, Guangdong, China\\
$^{9}$ Institute for Nuclear Research of Russian Academy of Sciences, 117312 Moscow, Russia\\
$^{10}$ School of Physical Science and Technology \&  School of Information Science and Technology, Southwest Jiaotong University, 610031 Chengdu, Sichuan, China\\
% ...

\end{center}

\vspace{10pt}

\begin{abstract}
Pulsar wind nebulae (PWNe) are bubbles of relativistic particles, powered by the rotational energy loss of the central pulsars. 
The Crab Nebula, powered by the Milky Way's most energetic pulsar, was discovered by the Large High Altitude Air Shower Observatory (LHAASO) as a PeV gamma-ray emitter, thereby establishing it as an extreme particle accelerator along with multiwavelength observations. Here we report LHAASO's detection of a point-like ultrahigh-energy (UHE, photon energy $E>100\,$TeV) gamma-ray source associated with the PWN powered by PSR~J1849-0001, a pulsar of spindown power 50 times lower than the Crab pulsar. The measured gamma-ray spectrum extends to PeV energies following a power-law distribution, with the PeV luminosity a few times higher than that of the Crab Nebula. Combined X-ray observations constrain the average magnetic field within the source to about $3\mu\,$G, and reveal an extreme particle acceleration efficiency approaching or even exceeding unity in the PWN, which we refer to as the ``Aquila Booster''. The result challenges the particle acceleration theory in PWN and implies non-ideal magnetohydrodynamics (MHD) conditions within the accelerator, potentially involving magnetic reconnection upstream of the termination shock.
\end{abstract}

The presence of the ``knee'' in the cosmic ray (CR) spectrum around 3\,PeV\cite{LHAASO2024_CRknee, KASCADE2005_CR, ASgamma2008_CR, IceCube2019_CR} ($1\,\rm PeV=10^{15}\,$eV) is considered as the limit of proton acceleration capacity for the majority of CR sources in the Milky Way. On the other hand, contributions from extra-Galactic CR accelerators start to dominate only above a few 1000\,PeV as indicated by the ``ankle'' structure in the CR spectrum. Therefore, a few Galactic factories of CRs must operate in extreme conditions and produce protons and nuclei well beyond multi-PeV energies, although identities of these super-PeVatrons are still a big puzzle. A clue to the mystery comes from detection of PeV gamma-ray photons. They are generated by multi-PeV electrons or tens of PeV protons, and hence indicate extreme particle acceleration capacity of their sources.

The Crab Nebula, a young and dynamic PWN, is a PeV gamma-ray emitter as revealed by LHAASO\cite{LHAASO2021_Crab}. LHAASO's measurement extended the spectrum of the Crab Nebula up to 1.1\,PeV, indicating a typical energy of $E_{\rm e}=2.3$\,PeV for the emitting electrons/positrons (hereafter, we do not distinguish positrons from electrons for simplicity). In general, the particle acceleration rate can be formulated as $\dot{E_e}=\eta eBc$ with $\eta$, $B$, $e$, and $c$ being the acceleration efficiency, the magnetic field, the elementary charge, and the speed of light respectively. The efficiency $\eta$ can also be given by $\eta=\mathcal{E}_{\rm eff}/B$, with $\mathcal{E}_{\rm eff}$ being the projection of the electric field averaged along a particle's trajectory, which is smaller than unity under the condition of classical electrodynamics and ideal MHD\cite{Aharonian2002_MHD}. The synchrotron radiation loss of electrons impose an upper limit on the maximum electron energy as $E_{\rm max}=5.8\,\eta^{1/2}(B/100\,\mu\rm G)^{-1/2}\,$PeV. To overcome the strong magnetic field of $B\sim 100\mu\,$G in the Crab Nebula, the inferred acceleration rate need reach at least 16\% of the theoretical limit\cite{LHAASO2021_Crab} (or $\eta\geq 0.16$) to explain the observation. Regardless of the specific particle acceleration mechanism, with such an extraordinary acceleration efficiency, protons can be energized to 10\,PeV scale without suffering radiative loss as electrons, as long as they are loaded in the particle acceleration zone of the Crab Nebula. 
However, even if the Crab Nebula is operating as an extremely efficient proton accelerator, its inferred PeV proton luminosity is not sufficient to explain the measured CR flux beyond the knee\cite{LiuRuoYu2021_Crab}. Also, it should be noted that the Crab pulsar is exceptional: it is less than 1,000 years old and has the highest spindown power ($4.5\times 10^{38}\,\rm erg~s^{-1}$) among all the pulsars detected in the Milky Way. As such, one may ask whether other less energetic PWNe could be extreme particle accelerators as well.

PWNe are the most numerous TeV--PeV gamma-ray sources detected in the Milky Way\cite{HESS2018_GalacticPlaneSurvey, HAWC2020_3Catalog, LHAASO2024_FirstCatalog}. The termination shock of the pulsar wind\cite{Lyubarsky2003, Sironi11, Arons12, Giacinti2018, LuYC21} formed by interactions between ultrarelativistic pulsar wind and the ambient medium has been suggested as a promising particle accelerator in PWN. 
In general, based on the Hillas criterion\cite{Hillas1984}, the maximum achievable particle energy is related to the spindown power of the pulsar $L_{\rm s}$ as\cite{Amato2021_ParticleAcceleration, Emma22} by $E_{\rm H}=e\sqrt{2\beta\varepsilon_{\rm B} L_{\rm s}/c}=7.8(\beta\varepsilon_{\rm B} L_s/10^{37}\rm erg~s^{-1})^{1/2}$\,PeV, where $\varepsilon_{\rm B}$ is the magnetic equipartition coefficient denoting the fraction of the spindown energy converted into the magnetic energy at the acceleration zone, and $\beta$ is the flow velocity in unit of $c$. We take $\beta=1$ hereafter as pulsar winds are ultrarelativistic. The practical maximum energy of accelerated particles can then be characterised by $E_{\rm max}=\eta E_{\rm H}$ from the perspective of the electric potential across the acceleration zone\cite{Aharonian2002_MHD}, which gives another constraint on the maximum energy in addition to the radiative-loss based constraint.

PSR~J1849-0001 is a fast-rotating pulsar in Aquila, and was initially discovered in the X-ray band\cite{Gotthelf2011_IGRJ18490-0000}. Its rotational period $P=38.5$\,ms is close to that of the Crab pulsar, but the spindown power $L_{\rm s} = 9.8 \times 10^{36}\,\rm erg~s^{-1}$ is about 50 times lower than the Crab pulsar. Based on the hydrogen column density obtained in X-ray spectral analysis of the pulsar, a distance of about 7\,kpc from the Earth is suggested\cite{Gotthelf2011_IGRJ18490-0000}. The pulsar is surrounded by a diffuse X-ray nebula\cite{Gotthelf2011_IGRJ18490-0000}. Recent analyses of Chandra X-ray observation of the region suggest that the nebula is composed of a compact bright inner nebula and a faint extended nebula\cite{2024ApJ...968...67G,Kim2024_PSRJ1849-0001}, although the latter does not show a sharp boundary. On the other hand, the High Energy Stereoscopic System (HESS) detected a faint extended TeV gamma-ray source HESS~J1849-000 in spatial coincidence with PSR~J1849-0001\cite{Terrier2008_IGRJ18490-0000, HESS2018_GalacticPlaneSurvey}, which is interpreted as the gamma-ray counterpart of the X-ray nebula. The extension of the HESS source is $0.09^{\circ}$ (or 11\,pc at a nominal distance of 7\,kpc), which is about three times of the extended X-ray nebula. 
The recent observation by the AS$_\gamma$ experiment has extended the spectrum of the source up to 320\,TeV\cite{Amenomori2023_ASgammaHESSJ1849-000}.

In the first catalog published by LHAASO\cite{LHAASO2024_FirstCatalog}, a point-like source, 1LHAASO J1848-0001u, is identified at an offset of 0.06$^\circ$ from the position of PSR J1849-0001 based on the data collected up to 2022 July 31. With the latest dataset of LHAASO up to 2024 July 31, a more precise measurement of the source is achieved and we update the source name as LHAASO~J1849-0002. The source significance reaches 34.0$\sigma$ above 25\,TeV and $23.7 \sigma$ above 100\,TeV. The best-fit position of the source above 25\,TeV (100\, TeV) is at $\rm R.A.=282.24^\circ \pm 0.01^\circ$ ($282.23^\circ \pm 0.02^\circ$) and $\rm Dec=-0.04^\circ \pm 0.01^\circ$ ($-0.05^\circ \pm 0.01^\circ$) presenting an offset of only $0.024^{\circ}$ ($0.039^\circ$) from the pulsar.
The 95\% C.L. upper limit above 25$~$TeV of the source extension ($r_{39}$) is $0.08^{\circ}$ (see Supplementary Information Section 2 for more details). In Figure~\ref{fig:sig_map}, we showcase the significance maps of this source in 2-40\,TeV, 25-100\,TeV and above 100\,TeV. The spectrum can be described with a broken power-law function, with spectral slopes being $\alpha_1 = 2.06\pm0.08$, $\alpha_2 = 3.08\pm0.10$ before and after the spectral break $E_{\rm br} = 66.5\pm7.9~\rm TeV$, as shown in Figure~\ref{fig:flux}. Adding an exponential cutoff at the high-energy end of the spectrum barely improves the spectral fitting, with the TS value increasing by only 2.2 (see Extended Data Figure~1). Given one additional free parameter with respect to the pure power-law function, we cannot claim the detection of the spectrum cutoff. The most energetic photon detected from LHAASO~J1849-0002 reaches 2\,PeV, with a positional deviation from the source by $0.08\pm 0.12^\circ$. The chance coincidence of the photon being originated from the cosmic-ray background and the diffuse gamma-ray background are only 0.02\% and 0.03\% respectively (see Methods and Supplementary Information Section 4 for more details). The energy of this photon is almost twice that of the highest energy of photons detected from the Crab Nebula. Since there is only one photon above PeV energy detected from this source, we combine the last three energy bins to get a more reliable estimation of the flux around PeV, finding a comparable flux to that of the Crab Nebula at PeV energies. The inferred PeV gamma-ray luminosity is even a few times higher than the Crab Nebula given a distance of 7\,kpc.

From the perspective of kinematics of the IC scattering process, electrons need be accelerated to at least 2\,PeV in this PWN to account for the 2\,PeV photon. The particle acceleration efficiency $\eta$ need be sufficiently high to overcome both the constraints from the radiative energy loss and the pulsar's spindown power. Following these two conditions, the requirement on $\eta$ is determined by the magnetic equipartition coefficient $\varepsilon_{\rm B}$ and the radius $R_{\rm acc}$ of the acceleration zone from the pulsar (see Methods). In Figure~\ref{fig:RB}, we show the least value of $\eta$ required to energize electrons to 2\,PeV for different combinations of $\varepsilon_{\rm B}$ and $R_{\rm acc}$. Even for an unrealistically high value of $\varepsilon_{\rm B}\to 1$, we need $\eta > 0.3$. This is already stronger than the constraint of $\eta$ in the Crab Nebula ($\eta>0.16$). We may impose a stronger constraint on $\eta$ if $\varepsilon_{\rm B}$ and $R_{\rm acc}$ can be better evaluated. 

Since the termination shock is a {widely discussed particle acceleration zone, its radius, denoted by $R_{\rm TS}$, may provide a reasonable estimation of $R_{\rm acc}$. 
Unlike in the Crab Nebula, the termination shock in the PWN of PSR~J1849-0001 is not resolved by X-ray observations, but it should be harbored within the compact inner PWN (see Extended Data Figures~2 and 3). The radius of the inner PWN $R_{\rm in}\approx 0.3\,$pc as revealed from X-ray observations sets an upper limit for the radius of the termination shock. From the theoretical point of view, the radius of the termination shock is determined by the pressure balance between the ram pressure of the pulsar wind and the pressure of the PWN immediately downstream of the termination shock\cite{GS2006}. 
The lower limit of the pressure in the PWN can be estimated via X-ray observations as detailed in Methods, which then sets up another upper limit of the pressure balance radius $R_{\rm bal}^*$. We then can take the smaller one between $R_{\rm in}$ and $R_{\rm bal}^*$ as the upper limit of termination shock radius $R_{\rm TS}^*$, as shown with the solid black line in Figure~\ref{fig:RB}. 
We see that the constraint on $\eta$ is mostly sensitive to $\varepsilon_{\rm B}$ if $R_{\rm acc}$ is not much smaller than $ R_{\rm TS}^*$. 
One way to estimate $\varepsilon_{\rm B}$ is through modeling the multiwavelength spectra of the PWN, including the X-ray flux measured by Chandra (see Extended Data Figure~4), and TeV -- PeV gamma-ray fluxes measured by HESS and LHAASO. Employing a simple one-zone leptonic model, in which the X-ray and gamma-ray fluxes are interpreted as the synchrotron radiation and IC radiation of electrons respectively (see the Extended Data Figure~5), the best-fit average magnetic field strength with the nebula is found to be $B = 2.81\pm 0.24~ \mu$G. It is weaker than the magnetic field in the Crab Nebula by about two orders of magnitude. Regarding the size of the source measured by HESS as the outer boundary of the PWN, we may obtain the total magnetic field energy in the PWN and translate it to $\varepsilon_{\rm B}\sim 10^{-3}$ at the termination shock, as long as the initial rotational period of the pulsar is not very close to the current period. A previous theoretical modeling of the SED and the X-ray intensity profile of this PWN using a more sophisticated multi-zone model results in a similar value of $\varepsilon_B \sim 0.004$ at the termination shock\cite{Kim2024_PSRJ1849-0001}. Indeed, it has been suggested that the magnetic energy of pulsar wind is converted to the kinetic energy far upstream of the termination shock\cite{Aharonian2012, Cerutti20}, leading to a small $\varepsilon_{\rm B}$ at the termination shock. Therefore, it would require $\eta \gg 1$ if electrons are accelerated to 2\,PeV at the termination shock.

This result has several profound implications. 
The condition $\eta \gg 1$ suggests that the local electric field in the termination shock (provided it being the acceleration zone of PeV electrons) is much stronger than that purely induced by the bulk motion of plasma (with a velocity $\beta$ in the unit of the speed of light) $\bm{\mathcal{E}}=-\bm{\beta} \times \bm{B}$. It is therefore challenging for particle acceleration models related to the first-order Fermi mechanism\cite{Rees1974, Giacche2017, Giacinti2018, Lyutikov2019}, at least in this PWN. On the other hand, the condition $\eta \gg 1$
indicates generation of strong resistive currents and violation of the ideal MHD condition.
This may be achieved in the striped wind scenario\cite{Coroniti90, Lyubarsky01}, in which magnetic reconnection events occur at the expense of the magnetic energy and energize particles via the reconnection electric field\cite{Lyubarsky08, Cerutti14, Sironi14}. It is important to note that, however, the termination shock is not the only candidate particle acceleration zone in PWNe. There have been suggestions of particle acceleration at radii either smaller or larger than the termination shock\cite{Aharonian2012, Cerutti20, Lou1996, Lyutikov2019}. For example, forward-reverse shock pairs may appear in pulsar wind given inhomogeneity in the pulsar wind, producing a highly compressed region with enhanced magnetic field at $R_{\rm acc}\sim 0.01$\,pc, and lead to particle acceleration\cite{Lou1996}. Besides, dissipation of the highly magnetized pulsar wind has also been suggested at a few tens of light-cylinder radii\cite{Aharonian2012, Cerutti20}. The stringent constraint on $\varepsilon_{\rm B}$ from multiwavelength observations may be circumvented if PeV electrons are accelerated at far upstream of the termination shock, where $\varepsilon_{\rm B}$ could be much higher and the extreme requirement of $\eta\gg 1$ may be relaxed. 
Nevertheless, even if $\varepsilon_{\rm B}\to 1$, the constraint from the available potential drop still demands $\eta>0.3$ (see Figure~\ref{fig:RB} and derivations in Methods), which is already notably higher than the $\eta>0.16$ inferred for the Crab Nebula. Furthermore, if PeV electrons are accelerated at a compact region far upstream of the termination shock with a high $\varepsilon_{\rm B}$, the synchrotron cooling becomes severe and the radiative-loss constraint would push $\eta$ above unity (e.g., the bottom-right part in Figure~\ref{fig:RB}). 
 
The extreme values of $\eta$ demanded by the LHAASO observations therefore pose a clear challenge to existing theoretical frameworks for PWNe. For instance, even in non-ideal MHD frameworks, fast reconnection requires the presence of turbulence\cite{Lazarian1999}. It consequently generates many small magnetic islands or tubes\cite{Lyubarsky08, Sironi14}, rather than a large-scale, coherent electric field over a sizable fraction of the accelerator. After averaging along a particle's trajectory, the effective electric field $\mathcal{E}_{\rm eff}$ and hence the acceleration efficiency $\eta$ could be substantially reduced. Thus, achieving $\eta$ values of order unity or higher in PWNe is highly nontrivial, calling for dedicated theoretical and numerical investigations.

In conclusion, LHAASO discovered that the gamma-ray spectrum of the Aquila Booster, the PWN powered by PSR J1849-0001, extends to PeV energies. The required minimum particle acceleration efficiency, $\eta\equiv \mathcal{E}_{\rm eff}/B>0.3$, exceeds that found in the Crab Nebula and may even surpass unity if the acceleration zone is not highly magnetized. It may imply a breakdown of the ideal MHD condition within the accelerator, which could reside far upstream of the termination shock. While in-depth theoretical study is needed to elucidate the specific mechanism accelerating electrons to PeV energies, this work suggests that extreme particle acceleration efficiency is not unique to the Crab Nebula and may be common in young PWNe. Although concrete evidence of high-energy proton acceleration in PWNe has not been found, given the exceptional acceleration capability, young PWNe might constitute a non-negligible population of PeV CR accelerators, provided that sufficient protons are loaded in pulsar winds.

\begin{figure}
\centering
\includegraphics[scale=0.37]{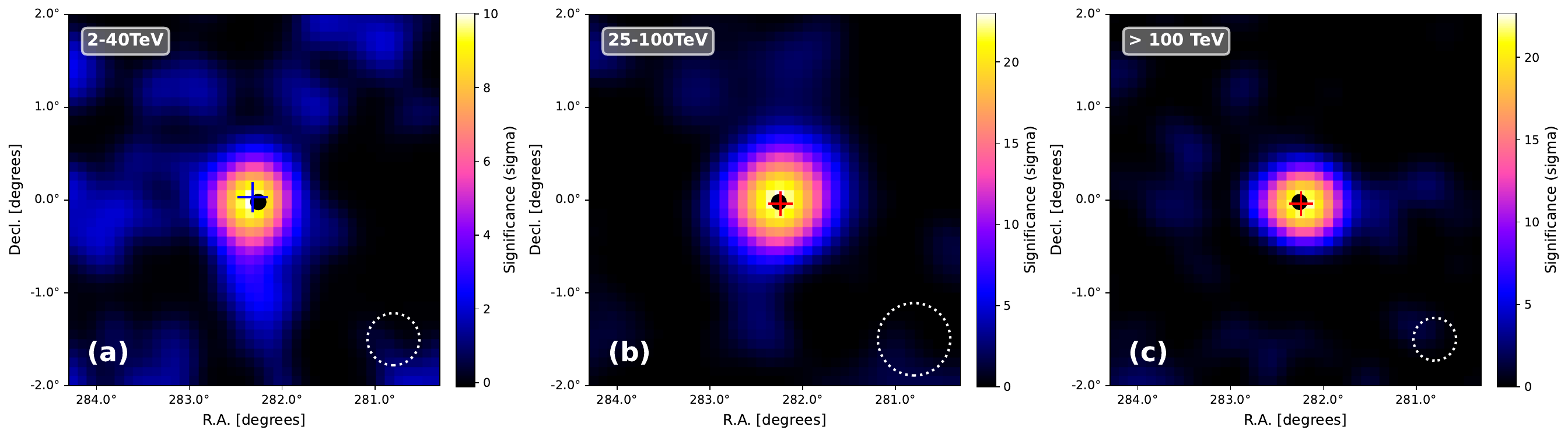}
\caption{\textbf{Significance maps of LHAASO~J1849-0002.} Panel (a): WCDA map with energy in $2-40~\rm{TeV}$. The blue plus sign indicates the best fit position. Panel (b): KM2A map with energy $25-100~\rm{TeV}$.
Panel (c): KM2A map with energy $> 100~\rm{TeV}$. The red plus signs in the panels (b) and (c) indicate the best-fit positions in these two energy bands. 
The black circle marks the positions of PSR~J1849-0001. The white circle in the bottom-right corner of the figures represent the instrument's PSF, indicating the $68\%$ containment radius.} 
\label{fig:sig_map}
\end{figure}

\begin{figure}
\centering
\includegraphics[width=0.5\textwidth]{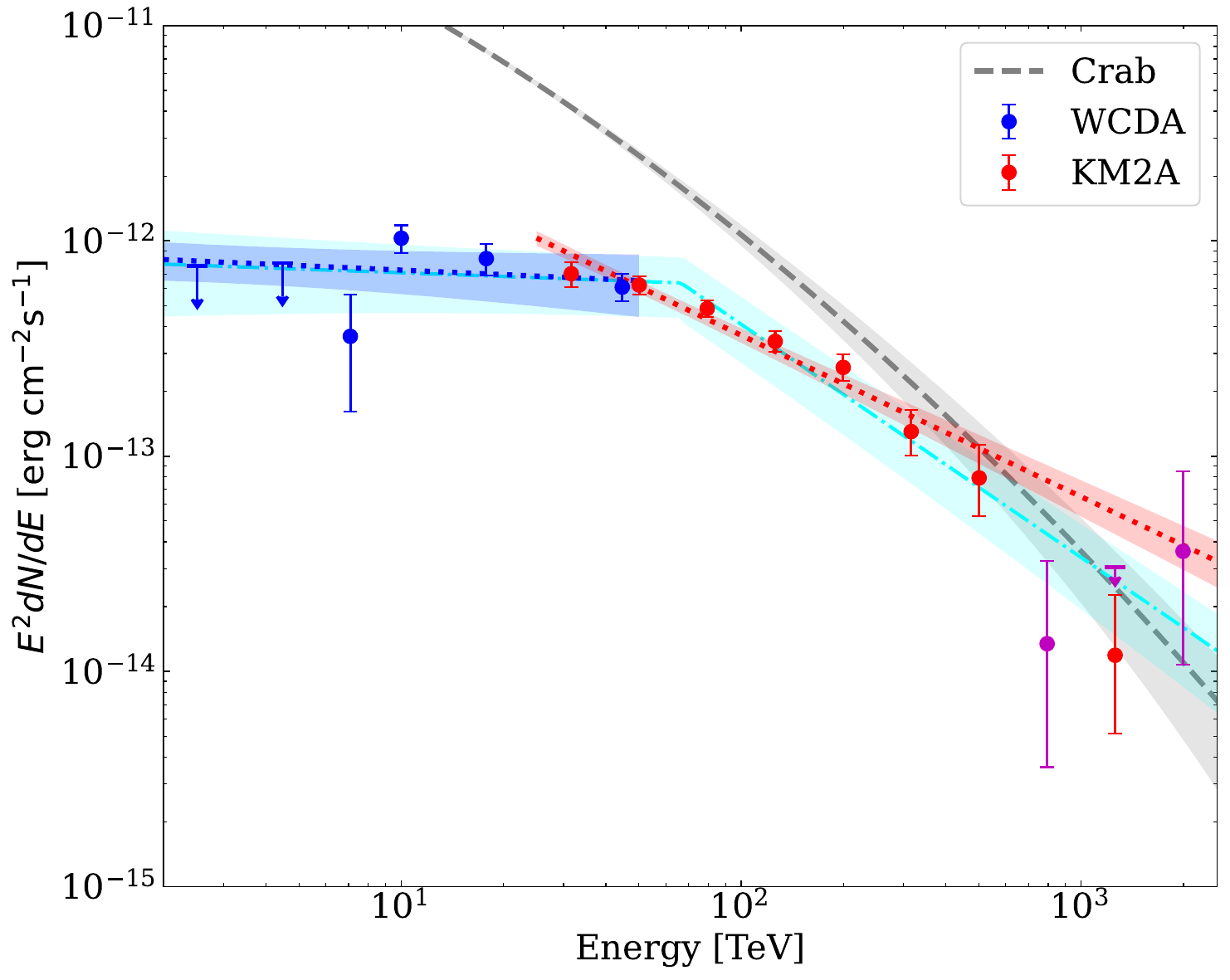}
\caption{\textbf{Gamma-ray spectrum of LHAASO~J1849-0002 measured by LHAASO.} Blue and red circles are flux measured by LHAASO-WCDA and LHAASO-KM2A respectively, where error bars represent $1\sigma$ uncertainties of the fluxes and arrows indicate $95\%$ upper limits. Blue and red lines represent the corresponding power-law function fitting results, and shaded region show the $1\sigma$ uncertainties of the fitting.
The cyan error band represents the overall spectrum described by the broken power-law function. The last red data point combines three energy bins from 630\,TeV to 2.5\,PeV, while the purple data points represent the individual data points for each of the three energy bins.
For comparison, the dashed grey curve shows the best-fit spectrum of Crab Nebula measured by LHAASO\cite{LHAASO2021_Crab}.}
\label{fig:flux}
\end{figure}

\begin{figure}
    \centering
    \includegraphics[width=0.6\textwidth]{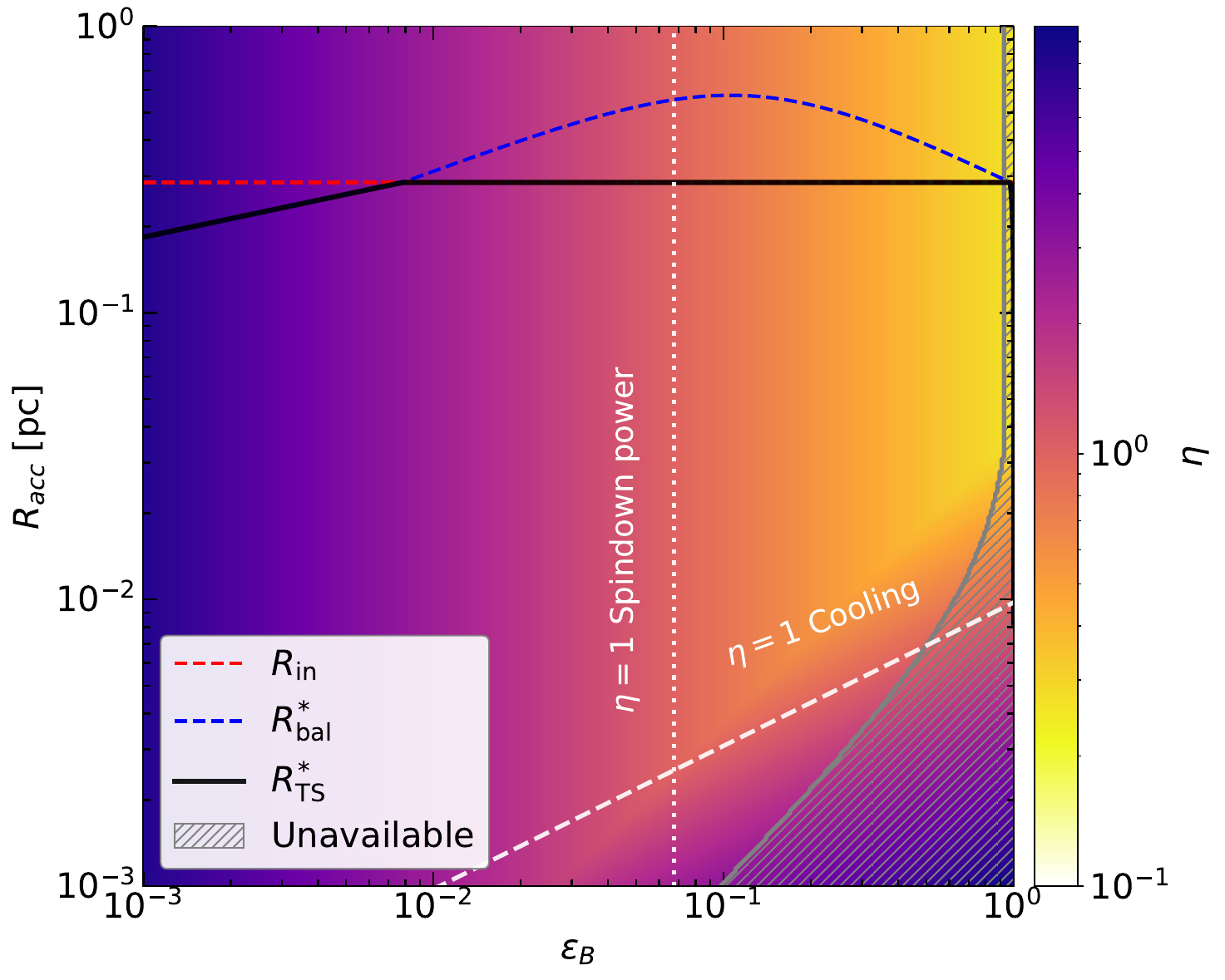}
    \caption{\textbf{The minimum particle acceleration efficiency $\eta$ required to accelerate 2\,PeV electrons under each combination of the radius ($R_{\rm acc}$) and the magnetic equipartition coefficient ($\varepsilon_{\rm B}$) at the acceleration zone.} Estimated upper limit of the radius of the termination shock, or $R_{\rm TS}^*=\min(R_{\rm in},R_{\rm bal}^*)$, as a function of $\varepsilon_{\rm B}$ is shown with the solid black line, where dashed red and blue lines represent the estimated upper limit of the pressure-balance radius $R_{\rm bal}^*$ and the radius of the inner PWN $R_{\rm in}$ based on the X-ray image, respectively. The dotted and dashed white lines mark where $\eta = 1$ is required, under the constraints from the spindown power of the pulsar and the radiative cooling, respectively. The grey-shaded area corresponds to $(\eta^2+1)\varepsilon_{\rm B} > 1$, where the combined electric and magnetic luminosity exceeds the spindown power of the pulsar, making this parameter region unavailable.}
    \label{fig:RB}
\end{figure}

\clearpage
\section*{Methods}
{\bf LHAASO Data Analysis}\\
In this study, both the data of the Water Cherenkov Detector Array (WCDA) and the Kilometer Square Array (KM2A) are used. For the KM2A, considering the energy resolution and statistics, one decade of energy is divided into five bins with a bin width of ${\rm log}_{10}E=0.2$. 
The sky in celestial coordinates is divided into grids with size of $0.1^{\circ} \times 0.1^{\circ}$ 
and filled with detected events according to their reconstructed arrival directions for each energy bin.
The maximum zenith angle of events was chosen to be $50^{\circ}$ in this paper. The gamma-ray data above $25~ \rm TeV$ used in this work were collected by the KM2A 1/2 array from December 27, 2019, to November 30, 2020, by the KM2A 3/4 array from December 1, 2020, to July 19, 2021, by the KM2A full array from July 20, 2021, to July 31, 2024, with a live time of 1570 days. For the WCDA, data used in this analysis were collected by full array from March 5, 2021 to July 31, 2024, with a live time of 1136 days. The number of triggered detector units $N_{\rm hit}$, which is selected as a shower energy estimator \cite{LHAASO2021ChPhC_WCDA}, is divided into six bands, i.e., $N_{\rm hit} \in (60-100, 100-200, 200-300, 300-500, 500-800, 800-2000)$. The reconstructed zenith angle is also less than $50^{\circ}$.
The background map is estimated by the direct integration method, which estimates the background using the events in the same directions in horizontal coordinate but different arrival times\cite{LHAASO_2021ChPhC}. 

In this study, we employed a 3D likelihood method for source modeling and analysis. The analysis was performed separately for WCDA and KM2A bands. The energy spectra of the sources were modeled with a power-law function, $dN/dE = N_0(E/E_0)^{-\alpha}$, and the parameters are evaluated using the forward-folding method. To obtain the flux point of a source in each energy bin, we fix the spectral index in each energy bin to the best-fit index obtained for the entire energy range, while the normalization factor of the spectrum in each energy bin are still set as a free parameter.  The flux point in each energy bin is calculated based on the spectral index and the normalization factor.  If TS value of an energy bin is smaller than 4, a flux upper limit with $95\%$ confidence level will be given by the likelihood profile method. The spatial models adopted either a two-dimensional(2D) symmetric Gaussian template or a point-like source model. A likelihood ratio test is performed to decide how many sources are needed to properly model in the region of interest (ROI) (see Supplementary Figure 2). The log-likelihood of the background-only model $\mathcal{L}_{b}$ and the log-likelihood $\mathcal{L}_{s}$ of the one-source model is computed, respectively. The test statistic (TS) defined as
\begin{equation}
    TS = -2(\ln\mathcal{L}_b-\ln\mathcal{L}_s)
\end{equation}
is used to compare the goodness of one model over the other. The TS-value is converted to a p-value, which is the probability of the data being consistent with the background-only hypothesis. TS follows a chi-square distribution, and the number of degrees of freedom is equal to the difference in the number of free parameters between the assumptions. 
In the case of a point source with fixed position, which has only one free parameter (the normalization of the spectrum, when ignoring the spectral distribution), the pretrial significance is $\sqrt{TS}$. The iteration process adds one source at a time to the fitting as long as the TS of N+1 sources is greater than that of N sources by more than 25. The procedure is iterated to determine the number of sources in the ROI. Our analysis additionally includes the effects of Galactic diffuse emission (GDE). The GDE resulting from the interaction of CRs with the ISM and background photons is an essential component of the gamma-ray sky, which has already been measured by LHAASO \cite{LHAASO2023_diffuse,LHAASO2025_diffuse}. Following the treatment in first LHAASO catalog\cite{LHAASO2024_FirstCatalog}, we used the distribution of the dust optical depth in the sky measured by Planck\cite{Planck2014}, which can be translated to the spatial distribution of the column density of gas, as the spatial template of GDE in the ROI. We assume the spectrum of GDE to be a power-law function in the energy range of KM2A, treating the normalization factor and spectral index as free parameters (the same for WCDA analysis, but with independent spectral parameters from KM2A), which will be obtained together with those of sources in the fitting.
With source spectrum model assumed as a simple power-law and spatial model as Gaussian extension, an iteration process based on multi-dimensional maximum likelihood is performed to derive the spectrum, position and extension.  
We identified 7 sources in this region, and the best-fit results are shown in Supplementary Table 1 and Supplementary Table 2. 
Notably, LHAASO J1849-0002 is spatially coincident with PSR J1849-0001. Consequently, we performed an extension test by comparing an extended Gaussian model with a point-like source model. The results showed that the Gaussian model provided no remarkable improvement to the fit, despite the additional degrees of freedom (more details see Supplementary Information Section 2).
\noindent As we mentioned above, the spectrum is assumed to be a single PL function for the data analysis of WCDA and KM2A separately (see Supplementary Information Section 3). The overall photon spectrum can is modeled with a broken PL function, i.e., 
\begin{equation}
  f(E) = A \times
    \begin{cases}
        \left(\frac{E}{E_{\rm br}}\right)^{-\alpha_1}, & E < E_{\rm br} \\
        \left(\frac{E}{E_{\rm br}}\right)^{-\alpha_2}, & E \geq E_{\rm br}
    \end{cases}
\end{equation}
$A$ is the normalization factor, $\alpha_1$ and $\alpha_2$ are the spectral indexes and $E_{\rm br}$ is the break energy. The best-fit results are $A = (9.15 \pm 2.90)\times 10^{-17} {\rm TeV^{-1} cm^{-2}s^{-1}}$, $\alpha_1 = 2.06 \pm 0.08$, $\alpha_2 = 3.08 \pm 0.11$ and $E_{\rm br} = 65.99 \pm 7.98\, \rm TeV$.
Note that for the highest-energy bins, there is only one photon detected in both the bin $[10^{2.8}, 10^{3.0}]\,$TeV and the bin $[10^{3.2}, 10^{3.4}]$\,TeV, while no photon is detected in the bin $[10^{3.0}, 10^{3.2}]$ TeV, with energy $E=654\pm 144\,$TeV and $E=2.02\pm 0.36\,$PeV. We also performed the fit by combining the three highest-energy bins into one. The resulting parameters ($A = (9.01\pm2.82)\times 10^{-17}~\rm TeV^{-1}cm^{-2}s^{-1}$, $\alpha_{1} = 2.06\pm0.08$, $\alpha_2 = 3.08\pm 0.10$ and $E_{\rm br} = 66.46\pm7.85~\rm TeV$) show no significant deviation from the original fit.
We further tested whether there is a clear cutoff at the high-energy end by comparing the goodness of fit between the power-law and exponentially cutoff power-law functions for spectrum above the 63\,TeV bin (at lower energies, the spectrum is subject to the break). Using the likelihood profile method, we searched for the best-fit cutoff energy, treating the normalization factor and spectral index as free parameters, as shown in the Extended Data Figure~1.  The best-fit cutoff energy is $E_{\rm cut}=602^{+\infty}_{-383} \rm TeV$. Since the model involves two free parameters, the upper error of the cutoff energy corresponds to a decrease in the TS value of 2.28 relative to the best-fit value, as indicated by the blue dashed line in Extended Data Figure~1. This suggests that the upper bound cannot be constrained. Besides, the improvement of the exponentially cutoff power-law over the power-law is only $\Delta \rm TS = 2.2$, so that the exponential cutoff power-law fit is not favored given one more free parameter ($\Delta \rm TS\geq 16$ is required to claim the opposite). Therefore, we conclude that the LHAASO data does not show an evidence of spectral cutoff at the highest energies. For more details of the LHAASO data analysis, please refer to the Supplemental Information.


\noindent {\bf X-ray Data Analysis}\\
In a PWN, if TeV -- PeV gamma-ray photons are generated through the IC radiation of electrons accelerated at the termination shock, these electrons are also expected to produce X-ray emissions via the synchrotron radiation. Therefore, the X-ray emission is supposed to have a comparable spatial extension to that of the gamma-ray emission. The size of the X-ray nebula of PSR~J1849-0001 reported in previous studies is smaller than that of the gamma-ray source reported by HESS\cite{Kim2024_PSRJ1849-0001,2024ApJ...968...67G}. This does not necessarily mean that the X-ray nebula has a more compact boundary. It could be simply due to that the X-ray emission becomes so dim and consequently unidentifiable at the outer region of the nebula. In other words, accelerated electrons are still producing X-ray synchrotron radiation outside the identified X-ray nebula but they are mixed with the background and not clearly seen.
As such, the local background subtraction region should be outside the gamma-ray source region, since otherwise the background flux could be overestimated given the contribution from relativistic electrons accelerated inside the PWN. We extracted the radial brightness profile and performed a spatially resolved spectral analysis using data from three Chandra Advanced CCD Imaging Spectrometer (ACIS) I-array observations (ObsIDs 23596, 24494, and 24495), each with a field of view of $16.8^\prime \times 16.8^\prime$. These observations started on 2022 September 15, 2021 September 26 and 2022 September 11 in Very Faint Timed Exposure mode, respectively. Data processing was carried out using the Chandra Interactive Analysis of Observations (CIAO\cite{2006SPIE.6270E..1VF}) software package version 4.16 with the Calibration Database (CALDB) version 4.11.0. We reprocessed the data using the \texttt{chandra\_repro} tool and obtained cleaned exposure time of 48.5\,ks, 29.7\,ks and 29.7\,ks after filtering for good time intervals. The $2-7\,$keV images were then merged using \texttt{merge\_obs} (\texttt{psfecf=0.393}, corresponds to the 1$\sigma$ integrated volume of a 2D Gaussian), and \texttt{wavdetect} (\texttt{ellsigma=3 and scales="1 2 4 8 16"}) was applied to the merged event for detecting point sources which would be masked in subsequent analyses, this resulted in 92 detected sources, with excision radii ranging from a few arcseconds in the inner region to roughly $10–20\,$arcseconds toward the edge of the field of view. We extracted the radial brightness profile centered on the pulsar using \texttt{dmextract}. The radial profile analysis revealed two components of the PWN in the previous ACIS-S (the maximum radius of the circle centered on the pulsar is about $250^{\prime\prime}$) study\cite{Kim2024_PSRJ1849-0001}: a compact core characterized by a Gaussian, i.e., \( C_1 \exp\left(-\frac{r^2}{2\sigma^2}\right) \), and a diffuse component described by an exponential function, i.e., \( C_2 \exp\left(-\frac{r}{l}\right) \). The background (after point sources being subtracted) is assumed to be homogeneous within the field of view (FoV). However, because the effective area of the instrument is not uniform across the FoV, this constant surface brightness of background is folded through the position-dependent instrumental response, resulting in the dashed gray line in Extended Data Figure~2. We follow this result to re-analysis the data, with taking the area outside the gamma-ray-emitting region (i.e., the HESS source and the upper limit of size of the LHAASO source) as the background region. The best-fit parameters, 
$\sigma = {8.3^{\prime\prime}}_{-0.7^{\prime\prime}}^{+0.8^{\prime\prime}}\,$ and \(l = {43.9^{\prime\prime}}_{-2.1^{\prime\prime}}^{+2.2^{\prime\prime}}\) (68\% confidence level) were obtained using a Markov Chain Monte Carlo (MCMC) analysis with an improved background estimation, as shown in Extended Data Figure~2. The results are consistent with previous studies\cite{2024ApJ...968...67G,Kim2024_PSRJ1849-0001}.

\noindent The compact inner nebula harbors the accelerator and its size can be considered as the upper limit of the radius of the acceleration zone. Using the profile resolved by the high-resolution Chandra image, we defined an annular region from a radius of $6.6^{\prime\prime}$ (where the pulsar's emission due to the PSF drops to half of the PWN's emission) to $12.5^{\prime\prime}$ (68\% containment radius or $1.51\,\sigma$ of the Gaussian component) to estimate the spectrum of the inner PWN. We also integrate the X-ray flux over the source region of HESS (i.e., a circular region with a radius of $0.09^\circ$), excluding the pulsar's emission, to evaluate the total X-ray flux of the PWN. 
Extended Data Figure~3 shows the spectral analysis regions as well as background selection. All regions are point source-free, and the best-fit flux distributions in $2-7$\,keV were derived, as shown in Extended Data Figure~4. For the spectral analysis, we first subtracted the instrumental background using `stowed' observations (which are not exposed to the sky), scaled to match the $9.5-12$\,keV counts rate for each ObsID, and then extracted the
spectrum and the ARF/RMF files for the background region shown in Extended Data Figure~3. We modeled the background spectrum with several components: an unabsorbed thermal component representing emission from the Local Hot Bubble, an absorbed thermal component representing emission from the Galactic ISM and an absorbed power-law
with photon index $\alpha \sim 1.45$ representing the unresolved background of cosmological sources. We then extracted the spectrum and ARF/RMF files for the source region; in the spectral fitting, we simultaneously fit the source and background spectra, modeling the source emission with an absorbed power-law, i.e., \( \frac{dN}{dE} = e^{-\sigma_{\rm ISM}(E)N_{\rm H}}N_0 \left({\frac{E}{E_0}}\right)^{-\Gamma} \), where $\sigma_{\rm ISM}$ is the photoionization cross section of the ISM and $N_H$ is the hydrogen column density. 
After grouping the source spectra to ensure at least 50 counts per bin for the inner PWN and for the HESS region, all spectra were fitted using the \texttt{Xspec} (v12.14.0h) in $2-7\,$keV and the hydrogen column density ${N_{\rm H}}$ of the source emission was fixed to $6.4\times10^{22}$ cm$^{-2}$\cite{Kim2024_PSRJ1849-0001} with the abundance model \texttt{wilm} and the ISM absorption model \texttt{tbabs} (if the \texttt{angr} abundance model is adopted, the ${N_H}$ can be fixed to $4.4\times10^{22}$ cm$^{-2}$\cite{Gotthelf2011_IGRJ18490-0000}) for a better constraint of the photon index. We obtained \( \Gamma = 1.59\pm0.35 \), \( N_0 = \left(1.92\pm0.98\right)\times10^{-5} \) photons/s/cm$^2$/keV for the inner PWN ($\chi^2$/d.o.f.=0.9) and \( \Gamma = 2.31\pm0.23 \), \( N_0 = \left(8.98\pm2.64\right)\times10^{-4} \) photons/s/cm$^2$/keV for the HESS region ($\chi^2$/d.o.f.=1.05) at $E_0=1\,$keV. 
The normalization of the inner PWN was scaled to the entire Gaussian based on the containment fraction of the two-dimensional Gaussian and finally we had a \( N_0 = \left(4.74\pm2.41\right)\times10^{-5} \) photons/s/cm$^2$/keV 
for the compact inner core. All these spectral parameters are reported with 1$\sigma$ uncertainties and the spectral analysis results were then used to fit the multi-wavelength SED.

\noindent {\bf Fermi-LAT Data Analysis}\\
The potential GeV $\gamma$-ray emission can give very useful constraints on the properties of CR injection from the pulsar and diffusion in the ISM. 
We analyzed Fermi-LAT data collected from 2008 August 4 to 2024 May 9 to search for extended gamma-ray emission associated with PSR J1849-0001. The dataset comprises Pass 8 (P8R3) SOURCE class events in the energy range of $10–500\,$GeV. A minimum energy threshold of 10 GeV was selected to suppress contamination from the pulsar's intrinsic emission, which typically exhibits a spectral cutoff at a few GeV. A square Region of Interest (ROI) of $20^\circ \times 20^\circ$ centered on the source was utilized, with a pixel size of 0.01$^\circ$. The analysis was performed using the Fermipy (v1.3.1) package and Fermitools (v2.2.0). We modeled the background using the standard Galactic diffuse emission ($gll\_iem\_v07$) and the isotropic component ($iso\_P8R3\_SOURCE\_V3\_v1$), while the initial source model was constructed based on the 4FGL-DR4 catalog. To reject Earth limb gamma-rays, a maximum zenith angle of $90^\circ$ was applied. During the fitting procedure, we freed the spectral parameters for all sources within a $4^\circ$ radius of the pulsar, along with the normalization factors of the Galactic and isotropic diffuse components. We noted that the closest neighbor, 4FGL J1848.2$-$0016, is offset from PSR J1849-0001 by $0.31^\circ$. We preclude a physical association between this source and the pulsar due to this spatial separation. The region surrounding the pulsar is otherwise devoid of known cataloged emitters.
Given the absence of cataloged sources in the vicinity, we employed the $find\_sources$ tool in Fermipy to search for potential new emission features with a significance $>4\sigma$. This procedure identified a gamma-ray excess (TS = 18.26) near the pulsar, which we tentatively designate as PS J1849.0-0000. The best-fit position of this excess is ($\rm RA = 282.263^{\circ}\pm0.020^{\circ}, Dec = -0.017^{\circ}\pm0.016^{\circ}$). Supplementary Figure 9 presents the residual TS map ($> 10 ~\rm GeV$) of the region, in which all background components—including cataloged point sources and diffuse emissions—have been modeled out to highlight the excess structure of PS J1849.0-0000. The TS value is defined as ${\rm TS} = 2(\ln\mathcal{L}_{1} - \ln\mathcal{L}_{0})$, where $\mathcal{L}_{1}$ and $\mathcal{L}_{0}$ represent the maximum likelihood values for models with and without the additional source, respectively. Our result is consistent with the findings of Xiao et al. (2024) \cite{Xiao2024}, who reported a similar excess with a TS of 19.48. However, since the significance does not reach the standard detection threshold of $5\sigma$ (TS $> 25$), we cannot firmly claim a detection and thus do not perform a detailed characterization of this source in the current work. Instead, to constrain the potential GeV emission from the PWN associated with PSR J1849$-$0001, we placed a hypothetical point-like source at the pulsar's position, modeled with a power-law spectrum (spectral index $\gamma = 2.0$). We then derived $95\%$ flux upper limits in five energy bins: 15–25, 25–40, 40–60, 60–110, and 110–200 GeV, using the standard likelihood method provided by Fermitools. During this bin-by-bin analysis, the spectral parameters of all background sources and diffuse components were fixed to the best-fit values obtained from the global analysis over the full energy range.

\noindent {\bf Multiwavelength Spectrum fitting}\\
We tried to fit the multiwavelength spectra of the source with the leptonic model using the Markov Chain Monte Carlo fitting routines of NAIMA, a package for the calculation of nonthermal emission from relativistic particles\cite{2015ICRC...34..922Z}. The spectrum of electron is also described as a broken power-law function as depicted above. The IC emission was calculated using background photon fields including the CMB, a far-IR component ($T = 30 \rm \,K$, $U = 2\times10^{-12}\, \rm erg/cm^3 $), a near-IR component ($T = 500\, \rm K$, $U = 4\times10^{-13} \, \rm erg/cm^3 $), and  a visible component ($T = 5000 \rm \, K$, $U = 1.5\times10^{-12}\, \rm erg/cm^3 $). 
By further considering the interstellar radiation field including the cosmic microwave background as the target radiation field of the IC process, we can search for the best-fit values of parameters via the Markov Chain Monte Carlo method. Note that, given that the pulsar is located at 7\,kpc away, a fraction of photons above 100\,TeV will be absorbed by the ISRF and CMB during the propagation. We have corrected for this effect in the energy spectrum fitting by multiplying a factor of $\exp(\tau_{\gamma\gamma})$ to the observed flux. $\tau_{\gamma\gamma}$ is the optical depth of gamma-ray photons due to $\gamma\gamma$ annihilation, which is given by\cite{Aharonian2004_book}
\begin{equation}
    \tau_{\gamma\gamma}=\int_0^{7\,\rm kpc}\int_0^{\infty} \sigma_{\gamma\gamma}(E,\omega)n_{\rm ph}(\omega;x;l,b)d\omega dx.
\end{equation}
Here, $\sigma_{\gamma\gamma}(E,\omega)$ is the angle-averaged (for isotropic radiation field) cross section of $\gamma\gamma$ annihilation between a gamma-ray photon of energy $E$ and a low-energy photon of energy $\omega$, which is analytically given by Aharonian (2004)\cite{Aharonian2004_book}. $n_{\rm ph}(\omega;x;l,b)$ is the differential number density of the background radiation field in the direction of Galactic coordinate ($l$, $b$) at a distance of $x$ from the Earth, which is based on the same ISRF model that we employed for calculating the IC radiation\cite{Popescu17}. The corrected energy spectrum is shown by blue circles in Supplementary Figure 10. The steeper slope in IC spectrum than in synchrotron spectrum after the spectral break is due to the Klein-Nishina effect. The derived parameters for the electron spectrum are
$s_1 = 2.34^{+0.04}_{-0.06}$, $s_2 = 3.29^{+0.41}_{-0.26}$, and $E_{\rm e, br} = 209.6^{+69.9}_{-60.9} \rm TeV$. The magnetic field was also treated as a free parameter in the fitting process, with the best-fit value being $B=2.81 \pm 0.24 \rm \, \mu G$. 
The total energy of electrons above 1\,TeV is found to be $5.0\times 10^{46}~\rm erg$. The best-fit result is shown by the black line in Extended Data Figure~5. 
We note that the difference between the spectral indexes before and after the break is roughly 1, implying that the break is caused by radiative cooling. Based on the best-fit magnetic field strength and the employed radiation field, we find that the cooling timescale of electrons at $E_{\rm e, br}$ is about 9\,kyr, which is much shorter than the characteristic age of the pulsar $\tau_{\rm c}=43.1\,$kyr. If the magnetic field strength and the radiation field density keeps constant during the evolution, the cooling timescale of electrons at the break energy would reflect the true age of the pulsar. If so, the related supernova remnant at this age could be visible over a wide range of energy band, which may be an interesting topic in the future. However, we should caveat that the magnetic field likely evolves with time in a PWN. At the early evolution stage when the nebula is smaller, the magnetic field is likely stronger. On the other hand, when the reverse shock of the supernova remnant crushes the PWN, the magnetic field may be enhanced greatly during the compression, but it could also drop remarkably during the later re-expansion stage of the PWN, according to dynamical evolution model of PWN\cite{Gelfand09}. The true age of the PWN would be younger than 9\,kyr if the magnetic field at earlier time is stronger, and vice versa. Given the weak magnetic field obtained above, it is more likely that the earlier magnetic field is stronger.

We may also estimate the equipartition coefficient $\varepsilon_{\rm B}$ at the termination shock based on the fitting result. The total energy of magnetic field inside the PWN is estimated to be $W_{\rm B}=(4/3)\pi R_{\rm PWN}^3 (B^2/8\pi)\approx 5\times 10^{46}(R_{\rm PWN}/11\,\rm pc)^3\,$erg. The total spindown power lost by the pulsar is given by $\Delta W_{\rm s}=L_{\rm s}\tau_{\rm c}[(P/P_0)^2-1]$ assuming a braking index of 3, with $P_0$ being the initial rotational period of the pulsar. The value of $P_0$ is unknown but a reasonable estimate would be between 10\,ms and 30\,ms, leading to $\Delta W_{\rm s}\sim 10^{49}-10^{50}\,$erg. We may estimate $\varepsilon_{\rm B}\sim W_{\rm B}/\Delta W_{\rm s}\sim 10^{-3}-10^{-2}$.

\noindent {\bf Estimation of the Termination Shock Radius}\\ 
The termination shock is widely suggested to be a potential particle acceleration site. Although its radius $R_{\rm TS}$ cannot be directly obtained with current observations, we may estimate an upper limit $R_{\rm TS}^*(\geq R_{\rm TS})$ as follows. 
Approximating the pressure in the shocked nebular interior by $P_{\rm PWN}^* = u_{\rm B}+ u_{\rm e}/3 $, we can calculate the pressure-balance radius $R_{\rm bal}^* = (L_{\rm s}/{4 \pi f_{\omega} c P_{\rm PWN}^*})^{1/2}$, where $u_{\rm B} = B^2/8 \pi$ is magnetic energy density (which is also the magnetic pressure) immediately downstream of the termination shock, and $u_{\rm e} = W_{\rm e,X}/V_{\rm PWN}$ is the energy density of X-ray emitting electrons with $W_{\rm e,X}$ being the total energy of X-ray emitting electrons. It can be estimated analytically by setting a pivot energy (e.g., at 2\,keV), so that we have $W_{\rm e,X}=f_{\rm e}L_{\rm X, 2keV}\tau_{\rm syn, 2keV}$. Here, $L_{\rm X, 2keV}=4\pi d^2F_{\rm X, 2keV}$ which is based on the observed flux $F_{\rm X, 2keV}\approx 10^{-13}\, \rm erg~cm^{-2}s^{-1}$ and $d=7\,$kpc as given by the X-ray observation. $\tau_{\rm syn, 2keV}$ is the synchrotron cooling timescale of electrons responsible for 2\,keV photon, i.e.,  $h\nu_{\rm c} = 3 h\gamma^2 eB/4\pi m_e c = 2\, \rm keV$, where $\gamma$ is the Lorentz factor of electrons and $\nu_c$ is the characteristic synchrotron frequency. Given $\tau_{\rm syn}=6\pi m_{e}^2c^3/(\sigma_{\rm T}E_{\rm e}B^2)$ with $\sigma_{\rm T}$ being the Thomson cross section, we obtain $\tau_{\rm syn,2keV}=4.4\times 10^{11}(B/3\mu{\rm G})^{-3/2}$\,s. $f_e \sim 1.1$ corrects for the total energy of electrons emitting over $2-7\,$keV in the inner PWN as observed by Chandra, which is evaluated numerically through fitting the observed X-ray spectrum of the inner PWN. $V_{\rm PWN}$ is the volume of the inner PWN region. For simplicity, we assume the region to be a sphere and get $V_{\rm PWN}=(4\pi/3)R_{\rm in}^{*3}$. Once the magnetic field is given, we may obtain $\tau_{\rm syn, 2keV}$, $P_{\rm PWN}^*$. Apparently, the particle pressure estimated via X-ray observations is an underestimation, because there are supposedly other contributions from lower-energy electrons including the thermal component, and relativistic protons which may be possibly loaded in the pulsar wind as well. In Figure~3, we assume the magnetic field immediately downstream of the termination shock to be the same as the magnetic field at the termination shock when calculating $R_{\rm bal}^*$. Finally, we obtain the upper limit of the termination shock radius by $R_{\rm TS}^*=\min(R_{\rm bal}^*, R_{\rm in})$.


\noindent {\bf Constraint on the Particle Acceleration Efficiency} \\
The spectrum of the PWN of PSR~J1849-0001 continues up to PeV energy, indicating that an extreme acceleration environment is required.
As already mentioned in the main text, the maximum electron energy achievable in PWNe is subject to two constraints. One is from the radiative loss of electrons assuming that the synchrotron radiation dominates. By equating the acceleration rate $\dot{E}_{\rm acc}=e\mathcal{E}_{\rm eff}c=\eta eB_{\rm acc}c$ and the synchrotron cooling rate $\dot{E}_{\rm syn}=(4/3)\sigma_{\rm T}c(E_e/m_ec^2)^2B^2/(8\pi)$, one may obtain $E_{\rm e,max}=5.8\,\eta^{1/2}(B/100\,\mu\rm G)^{-1/2}\,\rm PeV$, which is used for constraining $\eta$ in the Crab Nebula\cite{LHAASO2021_Crab}. If we employ the same equation and assume the magnetic field in the accelerator $B_{\rm acc}$ as the one obtained from SED modeling (i.e., $2.81\,\mu$G), we would obtain $\eta>0.003$. Indeed, the low magnetic field leads to an inefficient cooling and does not impose strong constraint on $\eta$ even considering additional cooling process by the IC radiation. However, the magnetic field obtained from the one-zone SED modeling should be understood as the volume-averaged field strength, and does not necessarily represents $B_{\rm acc}$. Without loss of generality, the magnetic field in the acceleration zone may be given by $B_{\rm acc}^2/8\pi=\varepsilon_{\rm B}L_{\rm s}/4\pi f_{\omega}R_{\rm acc}^2 c$, where $f_\omega$ is the equivalent filling factor for an isotropic pulsar wind. Then, the maximum electron energy limited by the synchrotron cooling can be reformulated as 
\begin{equation}\label{eq:cooling_epB}
    E_{\rm e, max}=6.3\,\eta^{1/2}f_\omega^{1/4}\left(\frac{R_{\rm acc}}{0.1\,\rm pc}\right)^{1/2}\left(\frac{\varepsilon_{\rm B}L_{\rm s}}{10^{37}\,\rm erg~s^{-1}}\right)^{-1/4}\,\rm PeV.
\end{equation}
The value of $f_\omega$ is usually assumed to be unity in literature\cite{GS2006}, and observations also suggested $f_\omega \lesssim 1$\cite{Ma2016}. Given the weak dependence of $\eta \propto f_\omega^{-1/2}$, we simply take $f_\omega$ to be 1 in Figure~3 noting that employing $f_\omega <1$ would only lead to more stringent requirement on $\eta$.  

The other constraint is from the electric potential drop across the acceleration zone. Although particles may make many irregular loops within the accelerator, the potential drop it may tap is $\mathcal{E}_{\rm eff}R_{\rm acc}$ or $\eta BR_{\rm acc}$ (where the constraint becomes equivalent to the Hillas criteria for $\eta=1$), and we may arrive at
\begin{equation}\label{eq:potential}
 E_{\rm e,max} = 7.8\,\eta \left(\frac{\epsilon_{\rm B} L_{\rm s}}{10^{37}\rm \,erg~s^{-1}}\right)^{1/2}\,\rm PeV.
\end{equation}
$f_\omega$ is canceled in this equation because the size of the acceleration zone is $\sim f_\omega^{1/2}R_{\rm acc}$. The true maximum electron energy is the smaller one between Eq.~\ref{eq:cooling_epB} and Eq.~\ref{eq:potential}, because electrons are subject to both constraints. Substituting $L_{\rm s}=9.8\times10^{36}~\rm erg~s^{-1}$ and $E_{\rm e,max}=2~\rm PeV$ into Eqs.~(\ref{eq:cooling_epB}) and (\ref{eq:potential}), we obtain $\eta=0.1\,\varepsilon_{\rm B}^{1/2}(R_{\rm acc}/0.1\rm pc)^{-1}$ and $\eta=0.26\epsilon_{\rm B}^{-1/2}$, respectively. For a given combination of $\epsilon_{\rm B}$ and $R_{\rm acc}$, the minimum $\eta$ is equal to the higher one between these two. Note that, if the electric field is close to or higher than the magnetic field, it would take a non-negligible fraction of the spindown power $\eta^2\varepsilon_{\rm B}$, so the condition $(\eta^2+1)\varepsilon_{\rm B}<1$ is needed, which leads to the grey-shaded region shown in Figure~3, and the smallest available value of $\eta$ is found to be about 0.3.

It may be also worth discussing the spindown of the pulsar. The evolution of the pulsar's spindown luminosity is generally given by\cite{Pacini1973} $L_s(t)=L_0/(1+t/\tau_0)^{(n+1)/(n-1)}$, where $L_0$ is the initial spindown luminosity, $\tau_0=2\tau_{\rm c}/(n-1)-t_{\rm age}$ is the initial spindown timescale and $n$ is the braking index. The true age of the pulsar $t_{\rm age}$ can be obtained if the initial rotation period $P_0$ and $n$ is given, via the formula\cite{Liu2020} 
\begin{equation}
    t_{\rm age}=\left\{
    \begin{array}{ll}
     2\tau_{\rm c}\left[1-(P_0/P)^{n-1} \right]/(n-1), ~~{\rm for}~~ n \neq 1    &  \\
     2\tau_{\rm c}\ln(P/P_0), ~~{\rm for}~~ n=1    & 
    \end{array}
    \right.
\end{equation}
If parent electrons of 2\,PeV photons are injected at earlier time when the spindown luminosity is higher, it would relax the requirement on $\eta$. On the other hand, the radiative cooling timescale of 2\,PeV electrons is only about 800\,yr in a magnetic field of $B=2.81\,\mu$G and the radiation field considered for SED modeling, so electrons injected more than 800 years ago would not survive today. We find the increase of the spindown luminosity of the pulsar at 800 years ago with respect to the present one is less than 5\%, as shown in Supplementary Figure 14. So we may conclude that considering electrons injected at earlier time will not change our conclusion notably.

\clearpage

\section*{Data availability}
The data supporting the conclusions of this paper are available at LHAASO web page (via the link http://english.ihep.cas.cn/lhaaso/) in the section ‘Public Data’ and the Chandra Data Archive
(https://cda.harvard.edu/chaser).

\section*{Code availability}
All the codes producing main results in this paper are developed by members of LHAASO Collaboration. These codes are available at the LHAASO web page (http://english.ihep.cas.cn/lhaaso/) in the section ‘Public Data’. The Fermi data processing code used in this work is publicly available at https://fermi.gsfc.nasa.gov/ssc/data/analysis/scitools. The Chandra data processing code used in this work is publicly available at https://cxc.cfa.harvard.edu/ciao.

\section*{Acknowledgement}
We would like to thank all staff members who work at the LHAASO site above 4400 meter above the sea level year round to maintain the detector and keep the water recycling system, electricity power supply and other components of the experiment operating smoothly. We are grateful to Chengdu Management Committee of Tianfu New Area for the constant financial support for research with LHAASO data. We appreciate the computing and data service support provided by the National High Energy Physics Data Center for the data analysis in this paper. This research has made use of data obtained from the Chandra Data Archive and the Chandra Source Catalog, and software provided by the Chandra X-ray Center 
(CXC) in the application packages CIAO. This research work is supported by the following grants: The National Natural Science Foundation of China No.12393852 (RYL), No.12393851 (ZC), No.12393853 (HZ), No.12393854 (RZY), No.12333006 (XYW), No.12121003 (JL), No.12205314(SW, No.12105301(HL), No.12305120 (XZ), No.12261160362 (ZL), No.12105294 (CL), No.U1931201 (QG), No.12375107 (LW), No.12173039 (XL), the Department of Science and Technology of Sichuan Province, China No.24NSFSC2319 (MZ), Project for Young Scientists in Basic Research of Chinese Academy of Sciences No.YSBR-061 (QY), and in Thailand by the National Science and Technology Development Agency (NSTDA) and the National Research Council of Thailand (NRCT) under the High-Potential Research Team Grant Program N42A650868 (DR).

\section*{Author Contribution}
R.Y. Liu initiated the project. R.Y. Liu, K. Wang and C.N. Tong led the writing of the manuscript. K. Wang, S.Z. Chen, and L.Y. Wang analysed the LHAASO data. S.C. Hu did the cross check of the LHAASO data analyses. C.N. Tong analysed the Chandra data. R.Y. Liu led the interpretation of the data. Z.Cao, F. Aharonian and D. Semikoz provided crucial suggestions. All the authors commented and edited the manuscript. 

\section*{Competing Interests Statement}
The authors declare no competing interests.

\section*{Correspondence}
Correspondence should be addressed to R.Y. Liu (ryliu@nju.edu.cn), K. Wang (k-wang@smail.nju.edu.cn), C.N Tong(cn\_tong@smail.nju.edu.cn), S.Z. Chen (chensz@ihep.ac.cn), and L.Y. Wang (wangly@ihep.ac.cn).

\clearpage

\clearpage

\vspace{10pt}   

\renewcommand{\figurename}{Extended Data Figure}
\setcounter{figure}{0}   
\renewcommand{\tablename}{Extended Data Table}
\setcounter{table}{0}

\begin{figure}[H]
\centering
\includegraphics[scale=0.36]{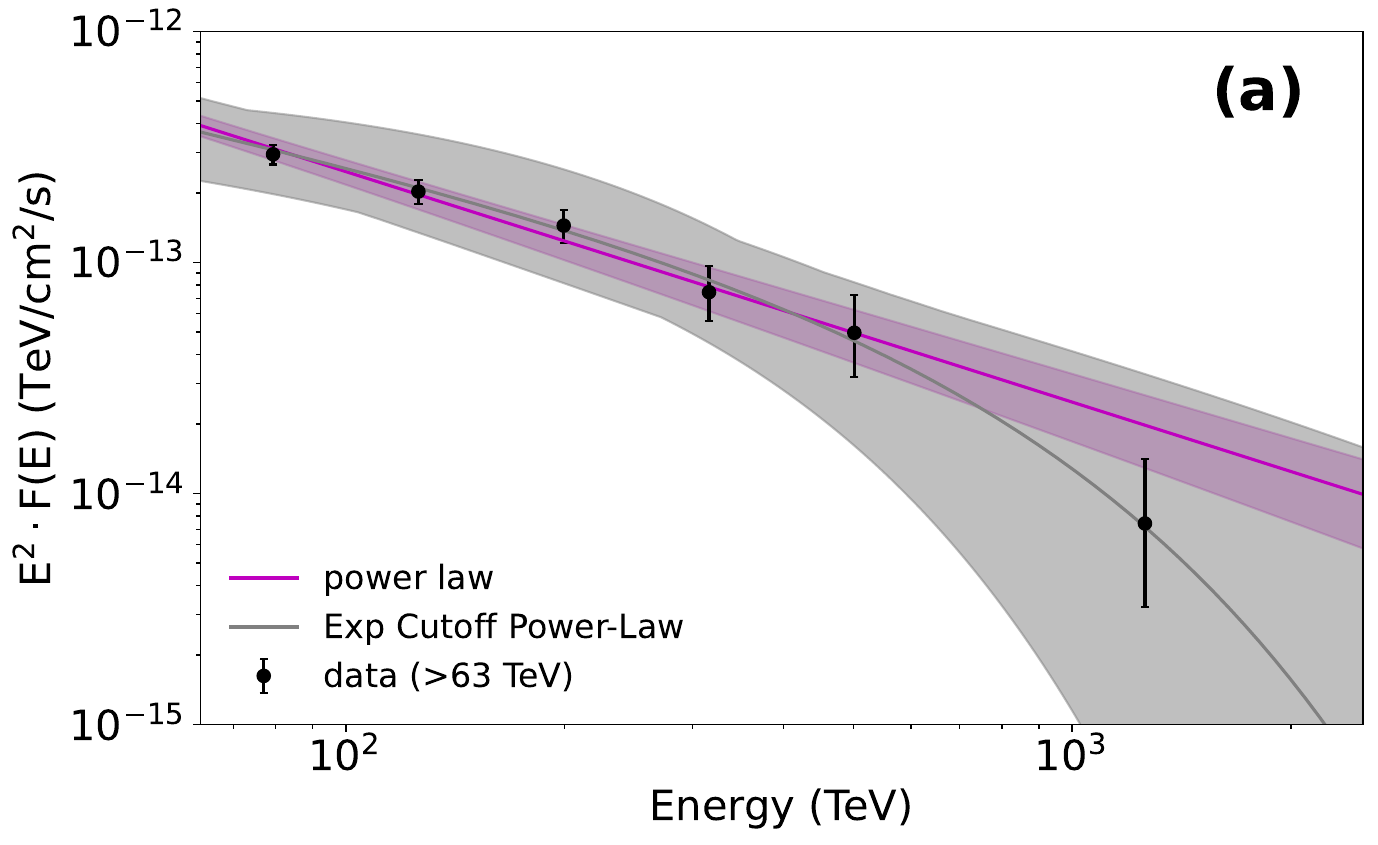}
\includegraphics[scale=0.445]{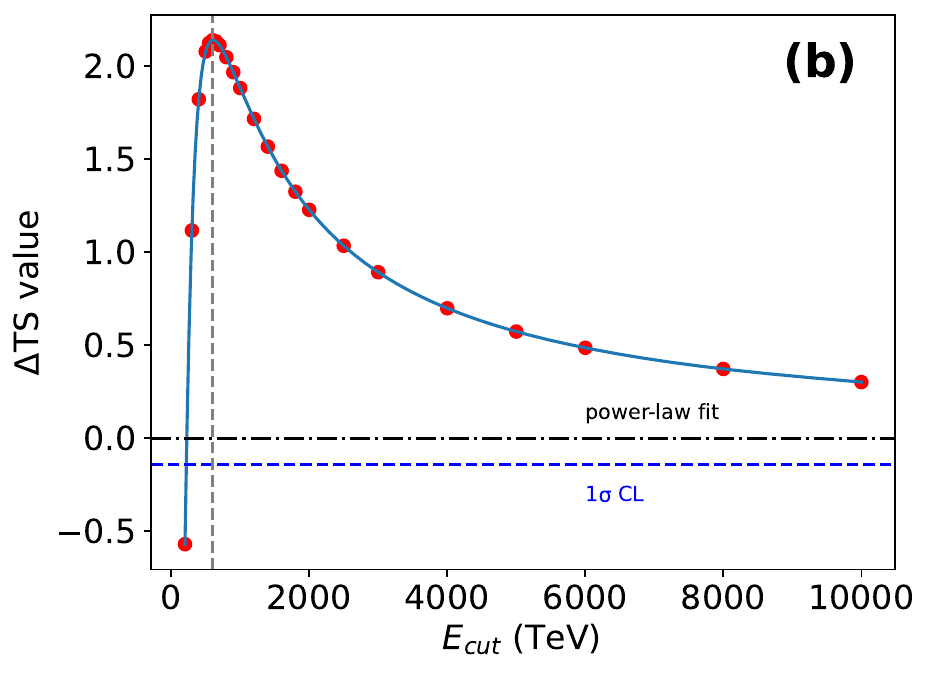}
\caption{Test of the high-energy cutoff energy of the spectrum. Left panel: Fitting to the spectrum of LHAASO~J1849-0002 above 63\,TeV. The black dots represent the data and its 1 $\sigma$ uncertainties. The magenta and grey lines represent the best-fit results with a PL and a PL exponential cutoff models, respectively. Shaded region marks the $1\sigma$ uncertain bands of the fit. Right panel: Values of $\Delta$TS as a function of $E_{\rm cut}$ under cutoff PL model with respect to PL model. The red dots represent the  
$E_{\rm cut}$ tested, and the blue solid line is the interpolation based on these points. The vertical grey dashed line indicates the $E_{\rm cut}$ corresponding to the maximum $\Delta$TS value. The horizontal black dot-dashed line indicates the TS baseline when fitting the spectrum with a power-law function. The blue dashed line represent the $1 \sigma$ C.L.}
\label{fig:SED_40TeV}
\end{figure}

\begin{figure}[H]
\centering
\includegraphics[width=0.7\textwidth]{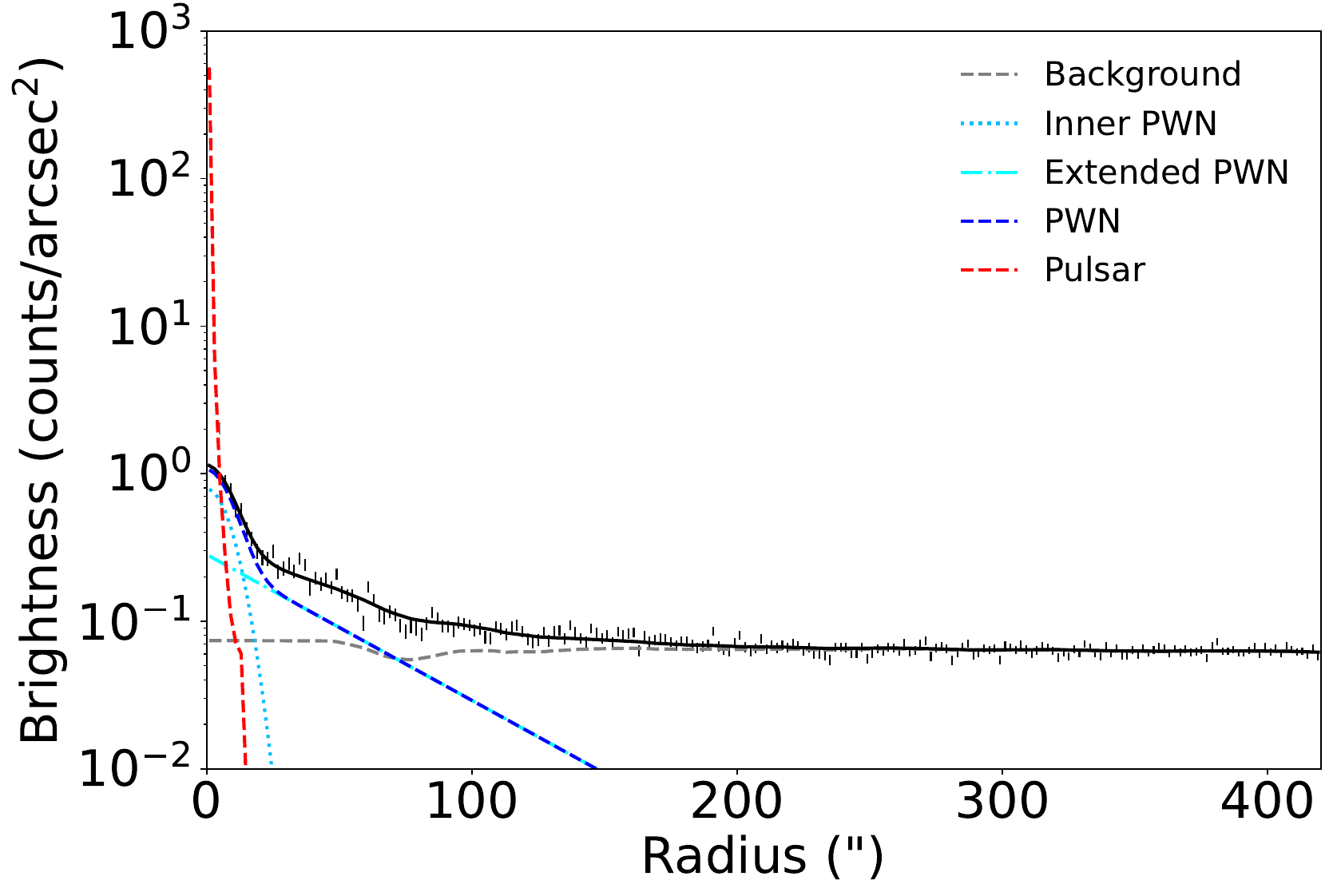}
\caption{Chandra radial profile in the $2-7\,$keV band, combining observations 23596, 24494, and 24495. Error bars of data points represent $1\sigma$ uncertainties. The dashed red, dotted deepskyblue, dashdot cyan, dashed blue and dashed grey represent the emission from the pulsar, inner PWN, extended PWN, PWN, and background, respectively, while the solid black curve shows the total emission from all models. The pulsar's emission was modeled using a MARX PSF simulation in CIAO.}
\label{fig:Chandra_profile}
\end{figure}

\begin{figure}[H]
\centering
\includegraphics[width=0.8\textwidth]{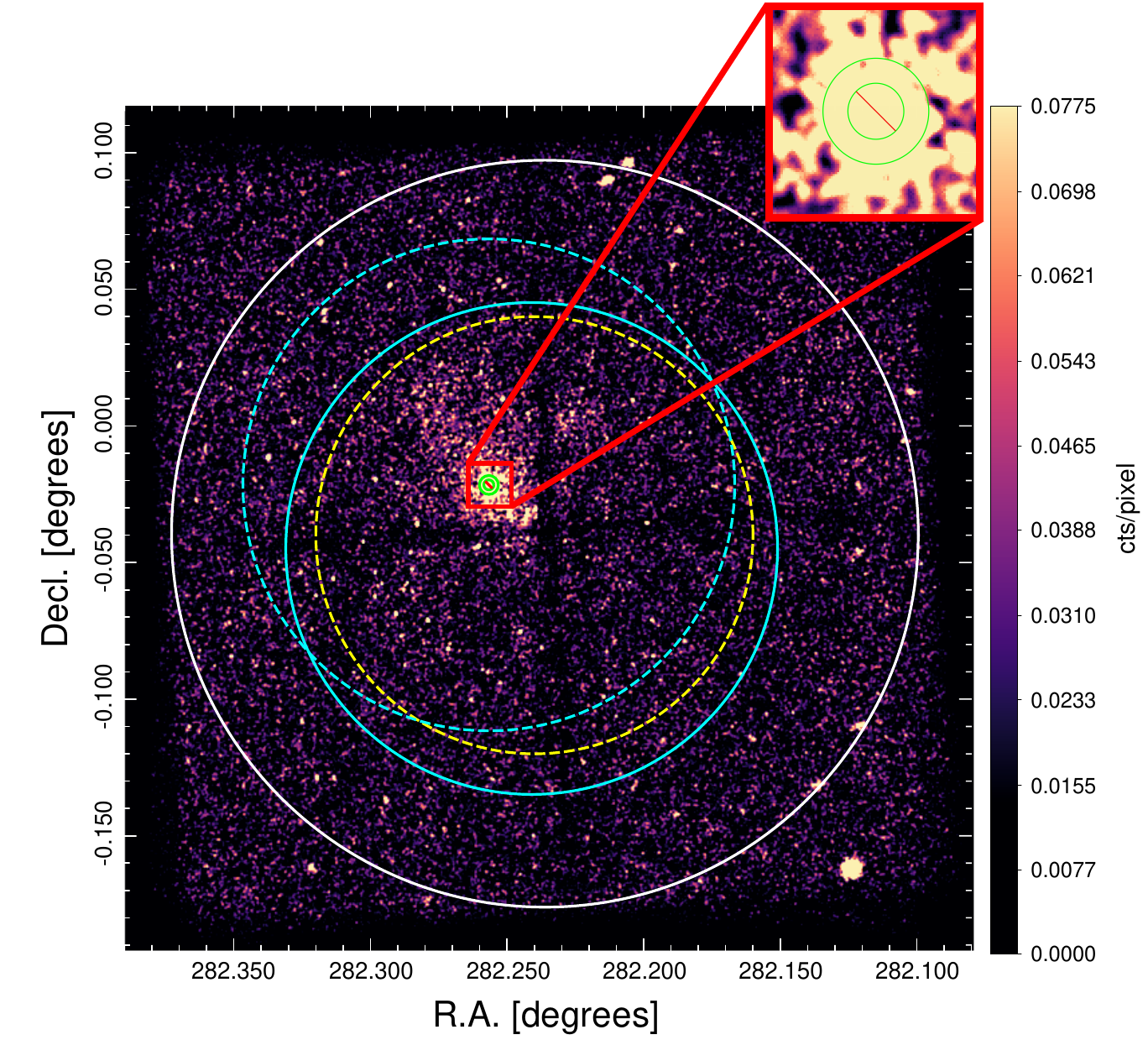}
\caption{Counts map of combined 3 Chandra observations in $2-7\,$keV, smoothed with a Gaussian kernel to 3$\sigma$ significance in linear scale and ZScale limits. The solid cyan circle indicates the HESS measurement region. The dashed cyan circle with a radius of $0.09^{\circ}$, centered at the position of the pulsar, defines the inner boundary for selecting a gamma-ray–free background region. The dashed yellow circle, marks the 95\% C.L. upper limit on the spatial extension of the LHAASO measurement. The background region is defined as the area within the white circle, excluding both the dashed cyan circle and the HESS source region. In the zoomed-in image, the region within the green annular ring represents the spectral extraction region for the inner PWN, with the central $6.6^{\prime\prime}$ circle excluded to minimize the contamination from the pulsar's emission.}
\label{fig:Chandra_skymap}
\end{figure}

\begin{figure}[H]
\centering
\includegraphics[width=0.8\textwidth]{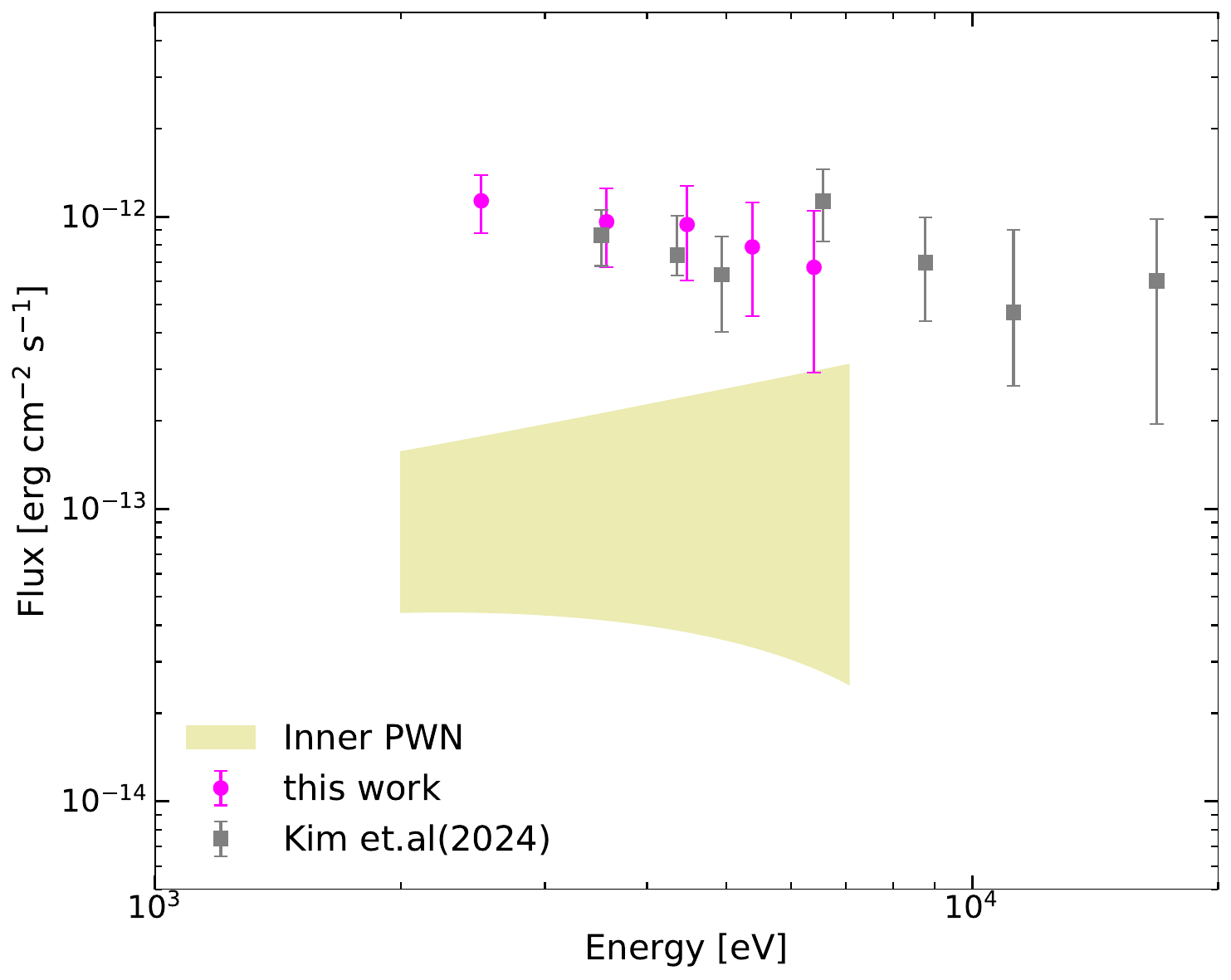}
\caption{X-ray spectra from different regions of the PWN. The magenta points show the X-ray flux integrated over the HESS region in this work, with the pulsar's emission excluded. The grey squares are measured within a circular region of radius $150^{\prime\prime}$ centered on the pulsar using Chandra and NuSTAR data, where the pulsar's contribution is minimized. The yellow band shows the X-ray flux of inner PWN obtained in this work, measured in an annulus with radii $6.6^{\prime\prime}$ to $12.5^{\prime\prime}$ centered on the pulsar, and scaled to the total Gaussian normalization. Error bars represent $1\sigma$ uncertainties.
} 
\label{fig:Xray_SED}
\end{figure}

\begin{figure}[H]
\centering
\includegraphics[width=0.8\textwidth]{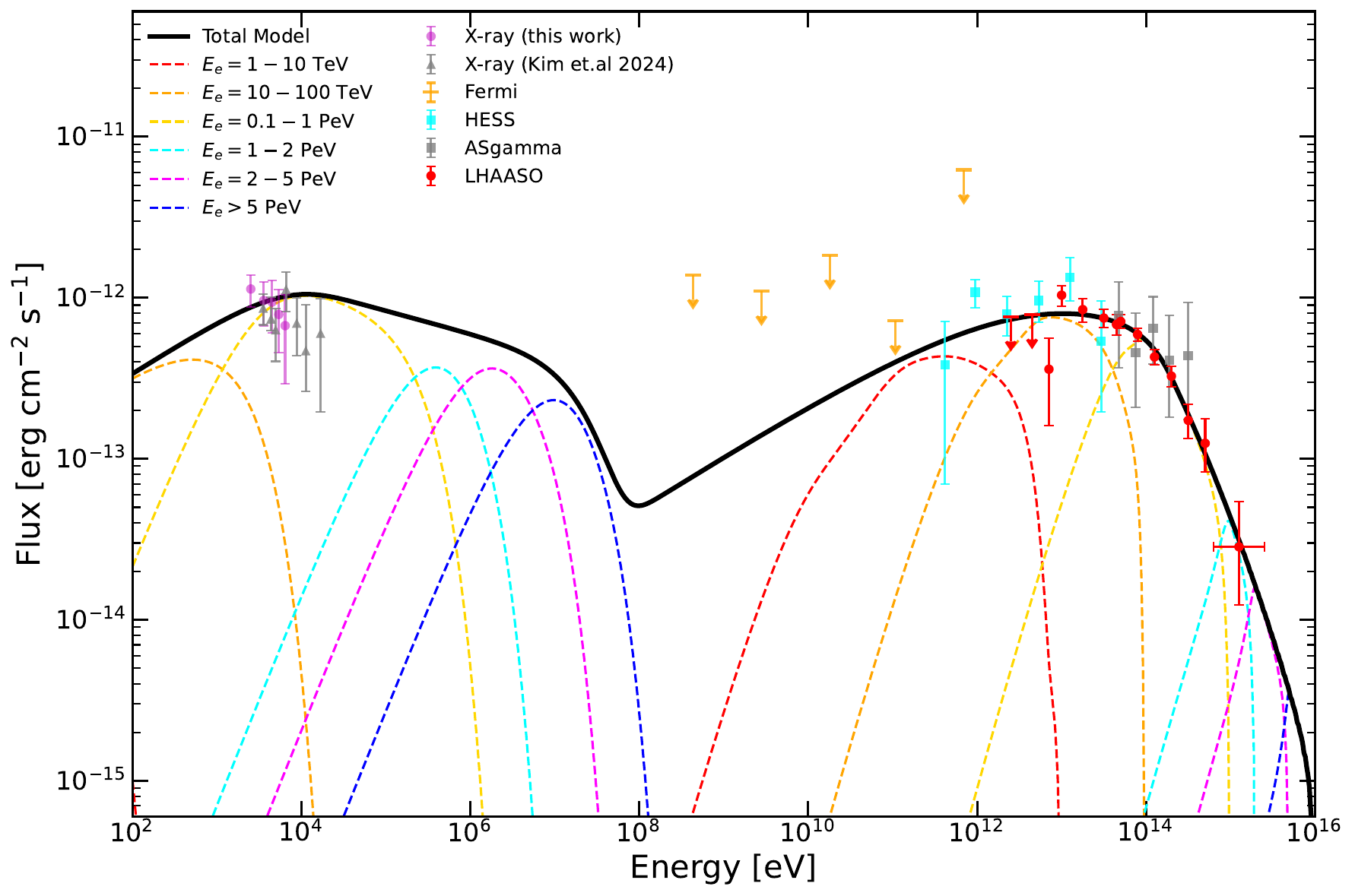}
\caption{Fitting to the multiwavelength spectrum of the PWN of PSR~J1849-0001 via synchrotron and IC radiations of electrons under the one-zone model. Dashed curves with different colors decompose contribution from different energies of electrons. The black curves represent the total flux of the synchrotron and IC components of radiation of electrons. The best-fit electron spectrum is a broken power-law function with the two indexes 
$s_1 = 2.34^{+0.04}_{-0.06}$, $s_2 = 3.29^{+0.41}_{-0.26}$, and the break energy $E_{\rm e, br} = 209.6^{+69.9}_{-60.9} \rm TeV$. The average magnetic field is $B = 2.81\pm0.24 \rm \mu G$. The red circles show the LHAASO measurement, noting that the last three energy bins are combined in this figure. Error bars represent $1\sigma$ uncertainties, and arrows indicate $95\%$ upper limits. Cyan squares show the measurement of HESS. Error bars show the $1\sigma$ flux uncertainties. Gamma-ray fluxes have been corrected for the absorption by ISRF and CMB, considering a source distance of 7\,kpc. The magenta points represents the X-ray flux measured by Chandra with $1\sigma$ uncertainties. Orange upper limits are GeV gamma-ray flux upper limits of 95\% confidence level based on the Fermi-LAT data.} 
\label{fig:SED_max}
\end{figure}

\clearpage

\bibliographystyle{naturemag}

\newpage
\onecolumn

\clearpage

\section*{Supplementary Information}
\addcontentsline{toc}{section}{Supplementary information}

\renewcommand{\figurename}{Supplementary Figure}
\renewcommand{\tablename}{Supplementary Table}

\renewcommand{\thefigure}{\arabic{figure}}
\renewcommand{\thetable}{\arabic{table}}
\setcounter{figure}{0}   
\setcounter{table}{0}   

\section{Systematic error of LHAASO data}
To estimate the systematic error associated with the instrument's pointing accuracy, we utilized the Crab Nebula as a standard calibration source.  We evaluated the position fitting accuracy independently for the WCDA and KM2A datasets by comparing the reconstructed positions with the precise reference coordinates of the Crab Nebula ($\rm RA = 83.63^{\circ}, Decl.  = 22.02^{\circ}$)\cite{LHAASO_2021ChPhC_sm}. We employed three comparative approaches to quantify pointing accuracy: First, we measured the best-fit positions of the TeV emission from the Crab Nebula using the full datasets from the two sub-arrays. The resulting coordinates were $(\rm{RA}, \rm{Decl.}) = (83.634^{\circ}\pm0.001^{\circ}, 22.014^{\circ}\pm0.001^{\circ})$ for WCDA and $(\rm{RA}, \rm{Decl.}) = (83.626^{\circ}\pm0.003^{\circ}, 22.018^{\circ}\pm0.003^{\circ})$ for KM2A.
Consequently, the angular separations between the fitted and reference positions were determined to be $\Delta \Psi \approx 0.007^{\circ}$ and $\Delta \Psi \approx 0.004^{\circ}$, for WCDA and KM2A respectively, where the separation is calculated as 
\begin{equation}\label{eq:separation}
\Delta \Psi = \sqrt{(\Delta \rm{RA} \cos \rm{Decl.})^2 + (\Delta  \rm{Decl.})^2}    
\end{equation}
Second, we examined the position offsets relative to the Crab Nebula in each energy bin. As shown in the left and middle panels of Supplementary Figure \ref{fig:crab_offset}, the deviations in RA and Decl. are presented across different energy bins, illustrating their distribution. By fitting these offsets, we obtained mean offsets of $(\Delta \rm{RA}, \Delta \rm{Decl.}) = (0.004^\circ, 0.006^\circ)$ for WCDA and $(\Delta \rm{RA}, \Delta \rm{Decl.}) = (0.004^\circ, 0.002^\circ)$ for KM2A. Substituting these mean values into the separation formula Eq.~(\ref{eq:separation}) yields a total offset of $\Delta \Psi = 0.007^\circ$ for WCDA and $\Delta \Psi = 0.004^\circ$ for KM2A.
Third, similar to the second approach, but we first calculated the total angular offset $\Delta \Psi$ at each individual energy bin based on Eq.~(\ref{eq:separation}), as illustrated in the right panel of Supplementary Figure \ref{fig:crab_offset}. We then fitted these total offset values to obtain a mean total offset of $\Delta \Psi = 0.007^\circ$ for WCDA and $\Delta \Psi = 0.004^\circ$ for KM2A. All three methods yielded consistent results.

\begin{figure}[H]
\centering
\includegraphics[scale=0.23]{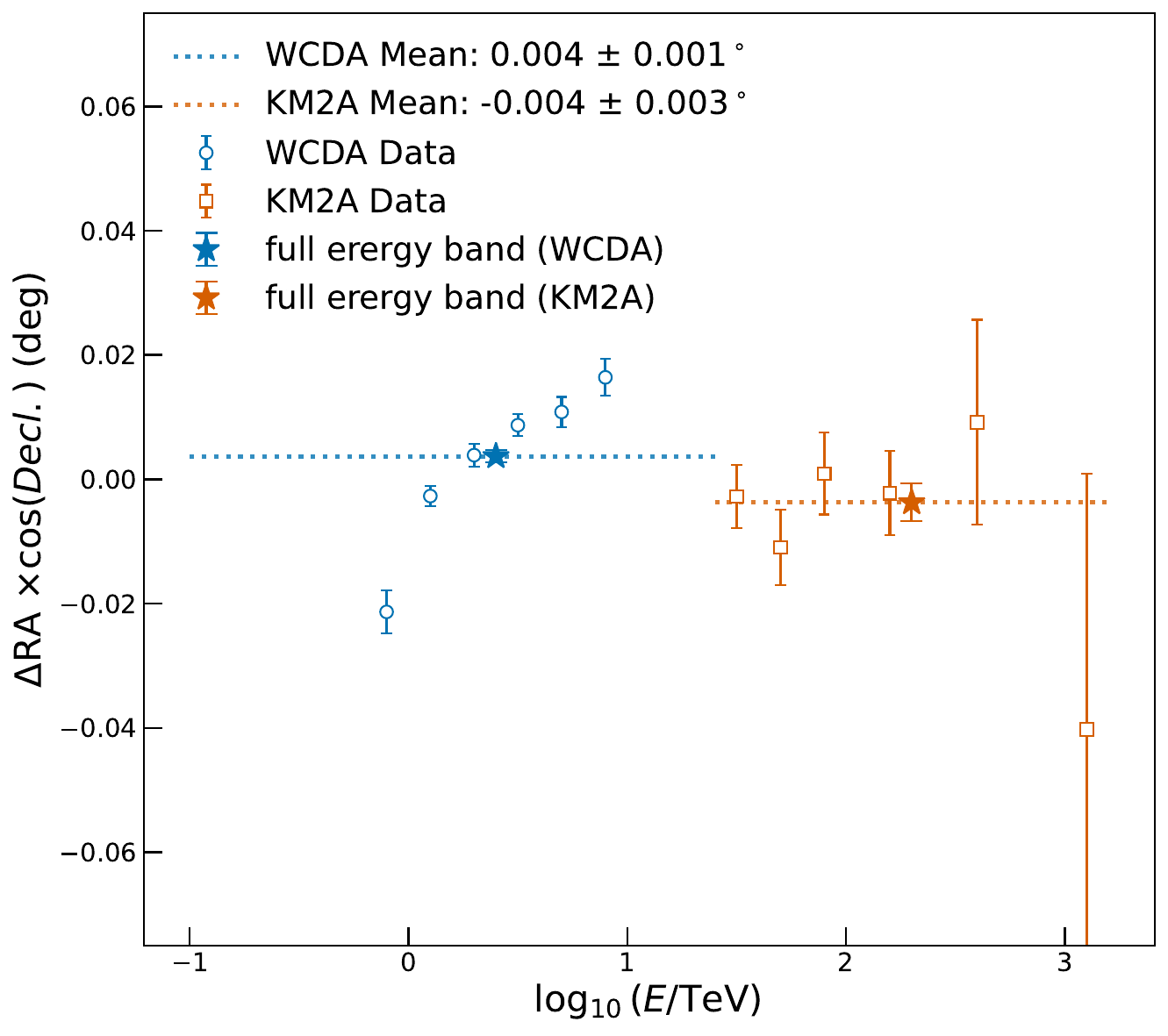}
\includegraphics[scale=0.23]{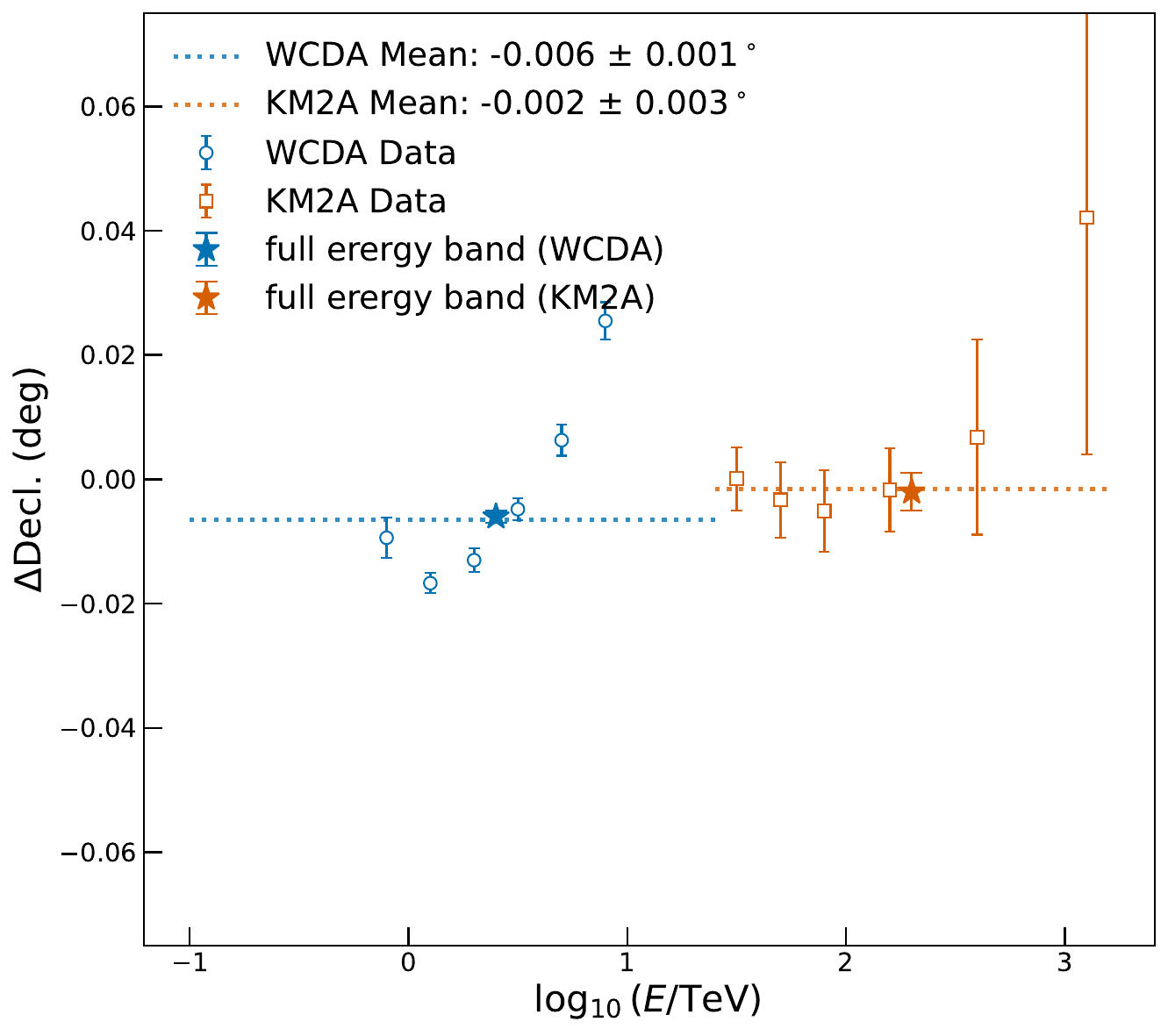}
\includegraphics[scale=0.243]{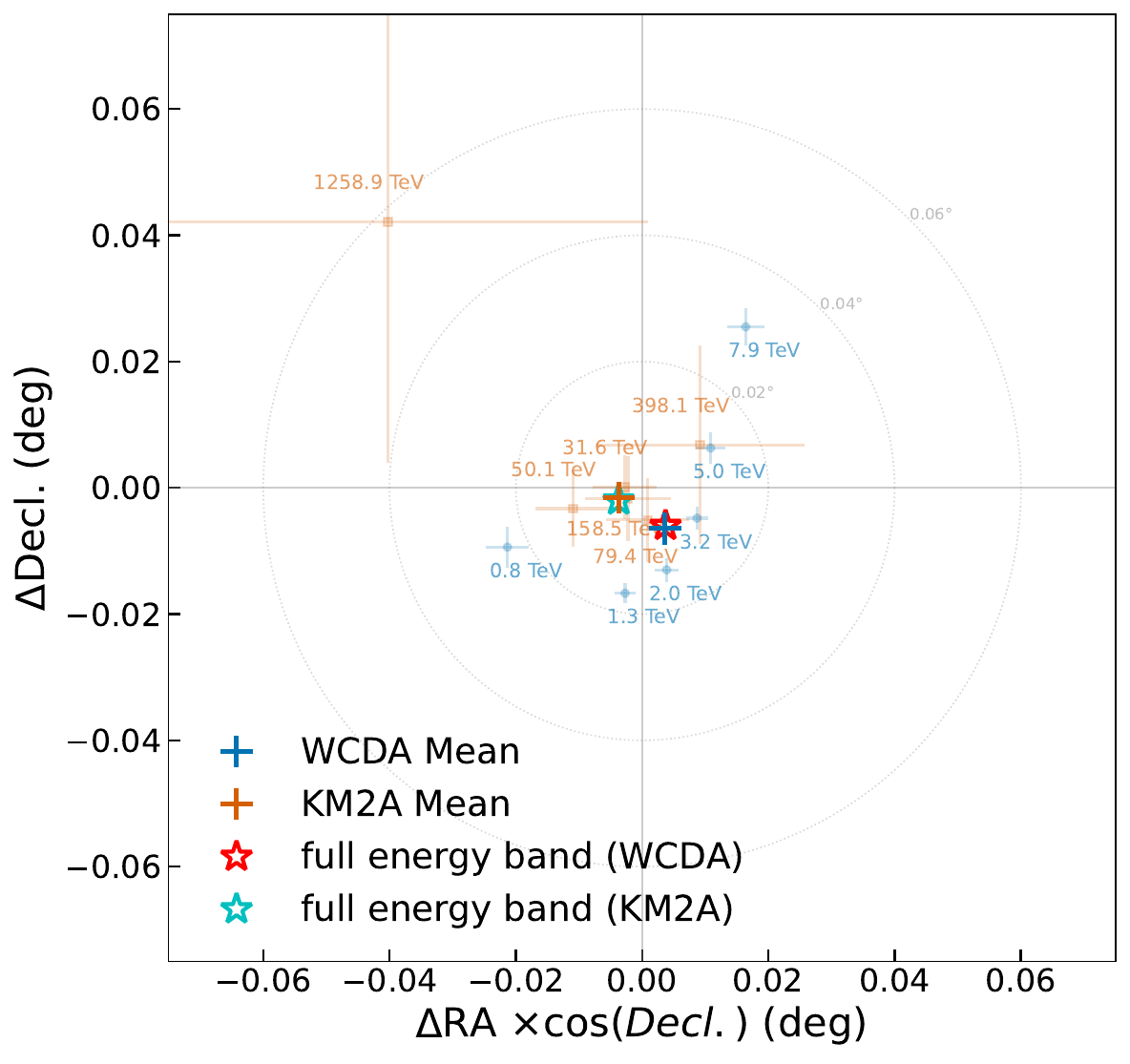}
\caption{The pointing error of LHAASO based on the position offset of Crab Nebula's emission measured by LHAASO to the nebula's reference position. The left and middle panels show the offsets in RA and Decl. relative to the Crab Nebula's position measured in different energy bins. Blue circles and orange squares represent the measurements from WCDA and KM2A in different energy bins. The blue dotted line and orange dotted line indicate the mean offsets for WCDA and KM2A, respectively. The blue star and orange star mark the best-fit position offsets derived from the full energy band data of WCDA and KM2A, respectively. Right panel: Total angular offsets across different energy bins. Numbers mark the median energy of each bin. The blue and orange crosses represent the mean values of these offsets for WCDA and KM2A, respectively. The red and cyan stars mark the best-fit positions offset obtained from the full energy band data of WCDA and KM2A.}
\label{fig:crab_offset}
\end{figure}

\section{Analysis of background sources}
We selected a $5^{\circ}\times5^{\circ}$ ROI, with RA in the range $[280.0^{\circ}, 285.0^{\circ}]$ and Decl. in the range $[-2^{\circ}, 3^{\circ}]$.
The significance map of this ROI is shown in Supplementary Figure \ref{fig:all_map}. 
The first LHAASO catalog shows that there are six sources in this area, numbered 1LHAASO J1848-0153u, 1LHAASO J1848-0001u (no WCDA detection), 1LHAASO J1850-0004u, 1LHAASO J1852+0052u, 1LHAASO J1857+0203u and 1LHAASO J1857+0245. Building on this, we identified seven sources in the region using an iterative approach. Each source is modeled with a 2D Gaussian template. In the fitting process, the influence of GDE was taken into account. Compared to the six sources model, the seven sources model improves the TS by 34.0 for WCDA and 25.1 for KM2A. LHAASO J1849-0002 (1LHAASO J1848-0001u) is associated with PSR J1849-0001. To study the significance of the extension of the source, we define ${\rm TS}_{\rm ext} = 2\rm{ln}(\mathcal{L}_{ext}/\mathcal{L}_{ps})$, i.e., twice the logarithm of the likelihood ratio of an extended source assumption to a point source assumption. We tested the spatial extension of the target source by comparing a point-like source with a Gaussian template, and found that the improvement from the Gaussian model over the point-like source is 1.1 for WCDA and 0.04 for KM2A. Therefore, we adopt a point-like source to describe the spatial morphology of the source. The best-fit results are listed in Supplementary Table \ref{tab:morfit_wcda} and Supplementary Table \ref{tab:morfit_km2a}. Although Supplementary Tables \ref{tab:morfit_wcda} and \ref{tab:morfit_km2a} list source names derived from the WCDA and KM2A best-fit positions respectively, we adopt the KM2A-based name as the standard throughout this work for consistency. We used the likelihood profile method to calculate the $95\%$ upper limit on the size of LHAASO J1849-0002, which is $0.11^{\circ}$ for WCDA and $0.08^\circ$ for KM2A. The likelihood profile distributions are shown in Supplementary Figure \ref{fig:Size_uplim}. In this work, the source extension is modeled using a symmetric 2D Gaussian spatial distribution, defined as $dN/d\Omega \propto \exp(-\theta^2 / 2\sigma^2)$. The reported source size corresponds to the standard deviation parameter, $\sigma$. Note that for a 2D Gaussian distribution, the radius $r = \sigma$ encloses approximately $39\%$ of the total flux. Therefore, the notation $r_{39}$ is used interchangeably with $\sigma$. Moreover, the radius containing $68\%$ of the flux ($r_{68}$) can be estimated as $r_{68} \approx 1.51 \sigma$. The upper limits of $0.11^\circ$ and $0.08^\circ$ refer to the $95\%$ confidence level upper limits on the parameter $r_{39}$.
Supplementary Figure \ref{fig:Remap} shows the residual map after subtracting all sources, with no significant excess remaining. Supplementary Figure \ref{fig:1D_Remap} shows the distribution of 1D residual significance at different energy ranges, which are in general in consistency with the normal distributions. In Supplementary Figures \ref{fig:all_map} and \ref{fig:Remap}, the best-fit position and size of the sources are marked by crosses and circles, with blue and red indicating WCDA and KM2A, respectively.

We searched for other potential high-energy counterparts near LHAASO J1849-0002, including high-luminosity pulsars ($>10^{34}~\rm erg/s$) and supernova remnants (SNRs). 
Supplementary Figure \ref{fig:all_map} identifies potential counterparts within a $1^{\circ}$ radius of  LHAASO J1849-0002 using magenta star symbols (filled: pulsars; unfilled: SNR Kes 78). With the exception of PSR J1849-0001, the closest of these objects is located $\sim 0.51^{\circ}$ away from the LHAASO J1849-0001. Given this significant spatial offset, they are unlikely to be the physical counterparts.

\begin{figure}[H]
\centering
\includegraphics[scale=0.3]{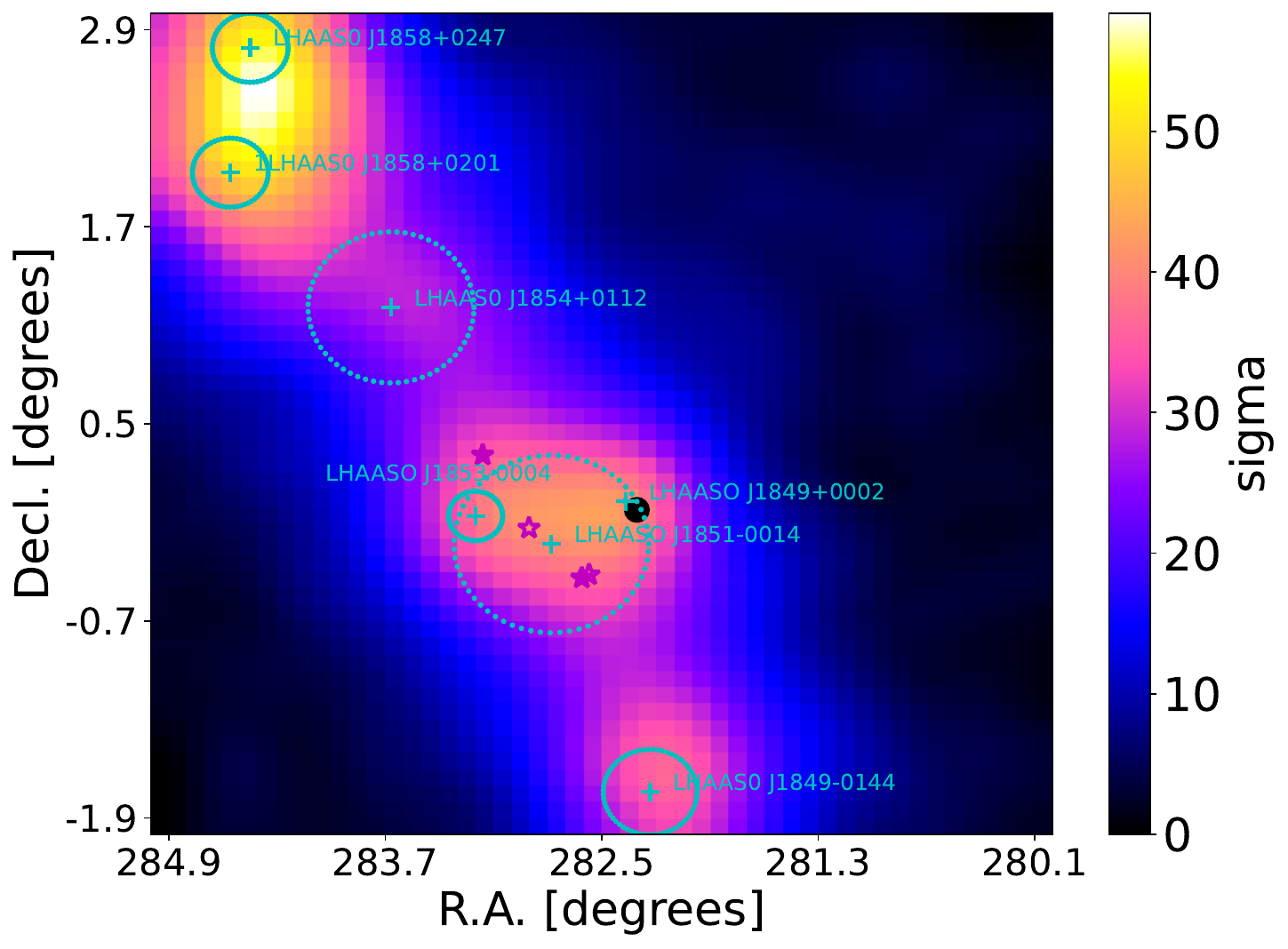}
\includegraphics[scale=0.3]{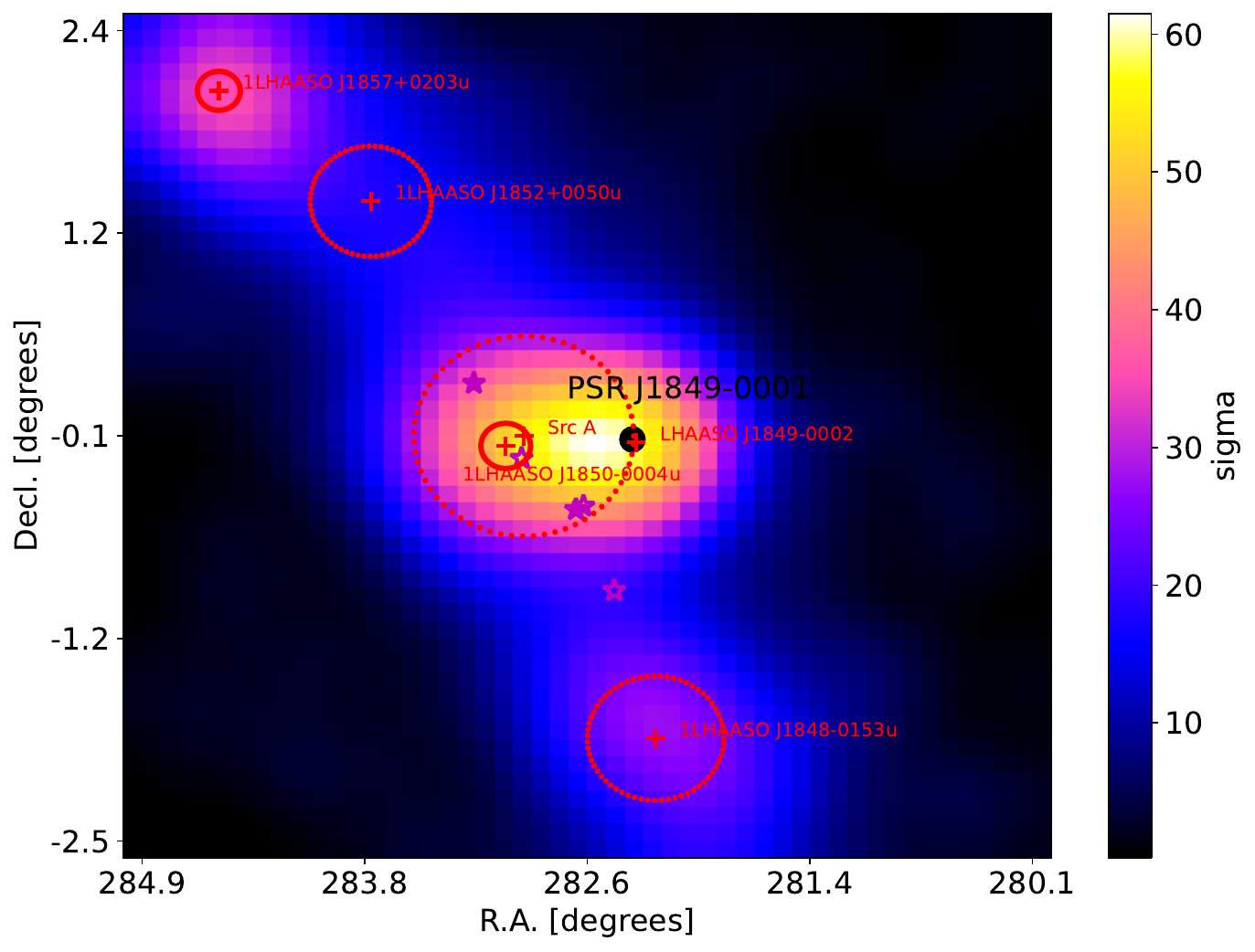}
\caption{Left panel: The significance map of ROI for WCDA ($2-40$\,TeV). Right panel: The significance map of ROI for KM2A ( $>25~\rm TeV$). The magenta star symbols in the figure represent other high-energy objects within $1^\circ$ of LHAASO~J1849-0002. Filled symbols indicate pulsars ($>10^{34}~\rm erg/s$), while hollow symbols indicate a SNR (Kes 78).} 
\label{fig:all_map}
\end{figure}

\begin{table}[htbp]
\centering
\caption{Best-fit model parameters of ROI for WCDA (2-40 TeV)}
\begin{threeparttable}
\begin{tabularx}{1\textwidth}{cccccc} 
\hline\hline 
\makecell{Name \\ LHAASO} & RA & Decl. & $r_{39}\tnote{a}$ & significance ($\sigma$) & \makecell{Counterpart \\ 1LHAASO} \\ 
\hline
J1849+0002 & $282.32^{\circ} \pm 0.02^{\circ}$ & $0.03^{\circ}\pm 0.02^{\circ}$ & - & 15.2 & J1848-0001u \\ 
J1853-0004 & $283.15^{\circ}\pm0.03^{\circ}$ & $-0.06^{\circ}\pm0.03^{\circ}$ & $0.15^{\circ}\pm0.05^{\circ}$ & 11.2 & J1850-0004u \\ 
J1851-0014 & $282.73^{\circ}\pm0.05^{\circ}$ & $-0.23^{\circ}\pm0.06^{\circ}$ & $0.54^{\circ}\pm0.03^{\circ}$ & 16.8 & new \\ 
J1854+0112 & $283.62^{\circ}\pm0.09^{\circ}$ & $1.21^{\circ}\pm0.08^{\circ}$ & $0.46^{\circ}\pm0.08^{\circ}$ & 12.4 &  J1852+0050u \\ 
J1858+0201 & $284.51^{\circ}\pm0.02^{\circ}$ & $2.03^{\circ}\pm0.03^{\circ}$ & $0.21^{\circ}\pm0.02^{\circ}$ & 28.5 &  J1857+0203u \\ 
J1858+0247 & $284.40^{\circ}\pm0.01^{\circ}$ & $2.79^{\circ}\pm0.03^{\circ}$ & $0.21^{\circ}\pm0.02^{\circ}$ & 38.1 &  J1857+0245u \\ 
J1849-0144 & $282.18^{\circ}\pm0.02^{\circ}$ & $-1.74^{\circ}\pm0.03^{\circ}$ & $0.26^{\circ}\pm0.02^{\circ}$ & 23.1 &  J1848-0153u \\ 
\hline\hline 
\end{tabularx}
\begin{tablenotes}
\footnotesize
\item[a] All extended sources in the table are modeled using a two-dimensional Gaussian model, with $r_{39}$ representing the best-fit value of the $39\%$ containment radius.
\end{tablenotes}
\end{threeparttable}
\label{tab:morfit_wcda}
\end{table}

\begin{table}[htbp]
\centering
\caption{Best-fit model parameters of ROI for KM2A ($>25~\rm TeV$)}
\begin{threeparttable}
\begin{tabularx}{1\textwidth}{cccccc} 
\hline\hline 
\makecell{Name \\ LHAASO} & RA & Decl. & $r_{39}\tnote{a}$ & significance ($\sigma$) & \makecell{Counterpart \\ 1LHAASO}\\ 
\hline
J1849-0002 & $282.24^{\circ}\pm0.01^{\circ}$ & $-0.04^{\circ}\pm0.01^{\circ}$ & - & 34.0 & J1848-0001u \\ 
J1852-0004 & $282.94^{\circ}\pm0.03^{\circ}$ & $-0.06^{\circ}\pm0.02^{\circ}$ & $0.15^{\circ}\pm0.04^{\circ}$ & 23.5 & J1850-0004u \\ 
J1851-0000 & $282.80^{\circ}\pm0.08^{\circ}$ & $-0.00^{\circ}\pm0.01^{\circ}$ & $0.67^{\circ}\pm0.10^{\circ}$ & 16.2 & new \\ 
J1855+0125 & $283.67^{\circ}\pm0.12^{\circ}$ & $1.42^{\circ}\pm0.10^{\circ}$ & $0.39^{\circ}\pm0.08^{\circ}$ & 9.4 &  J1852+0050u \\ 
J1858+0201 & $284.49^{\circ}\pm0.02^{\circ}$ & $2.01^{\circ}\pm0.02^{\circ}$ & $0.10^{\circ}\pm0.03^{\circ}$ & 24.7 &  J1857+0203u \\ 
J1858+0247 & $284.47^{\circ}\pm0.05^{\circ}$ & $2.79^{\circ}\pm0.08^{\circ}$ & $0.21^{\circ}\pm0.07^{\circ}$ & 10.9 &  J1857+0245u \\ 
J1849-0143 & $282.19^{\circ}\pm0.04^{\circ}$ & $-1.71^{\circ}\pm0.06^{\circ}$ & $0.30^{\circ}\pm0.05^{\circ}$ & 19.1 &  J1848-0153u \\ 
\hline\hline 
\end{tabularx}
\end{threeparttable}
\label{tab:morfit_km2a}
\end{table}

\begin{figure}[H]
\centering
\includegraphics[scale=0.4]{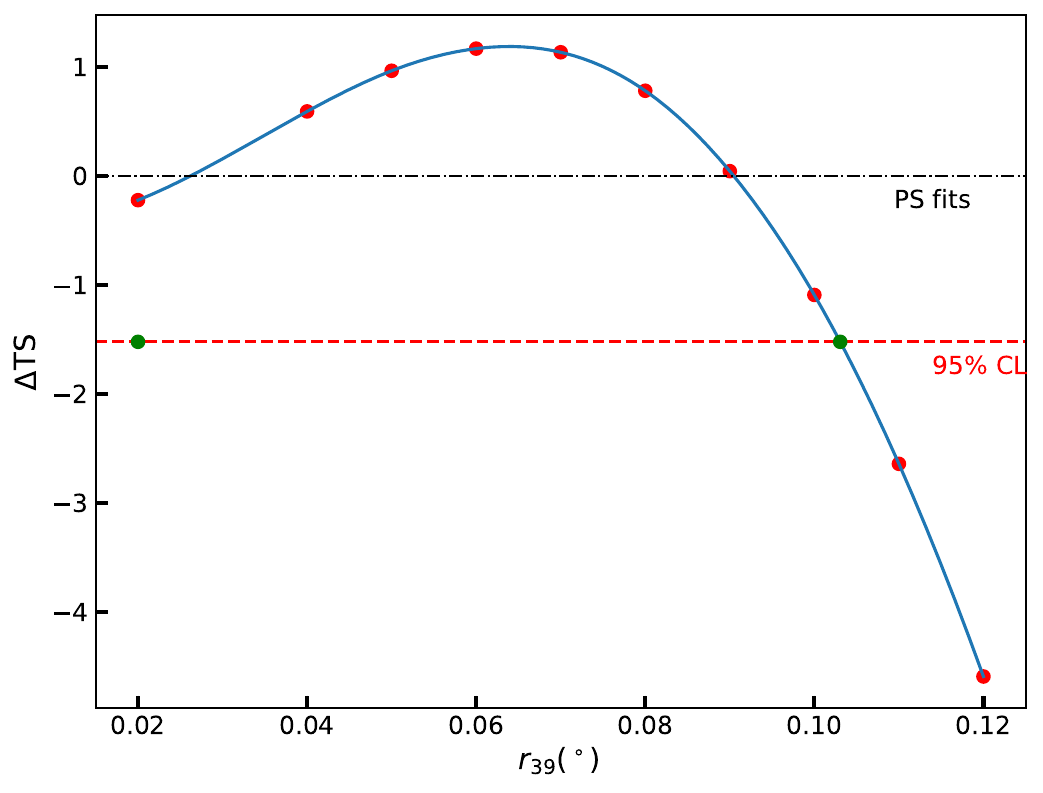}
\includegraphics[scale=0.4]{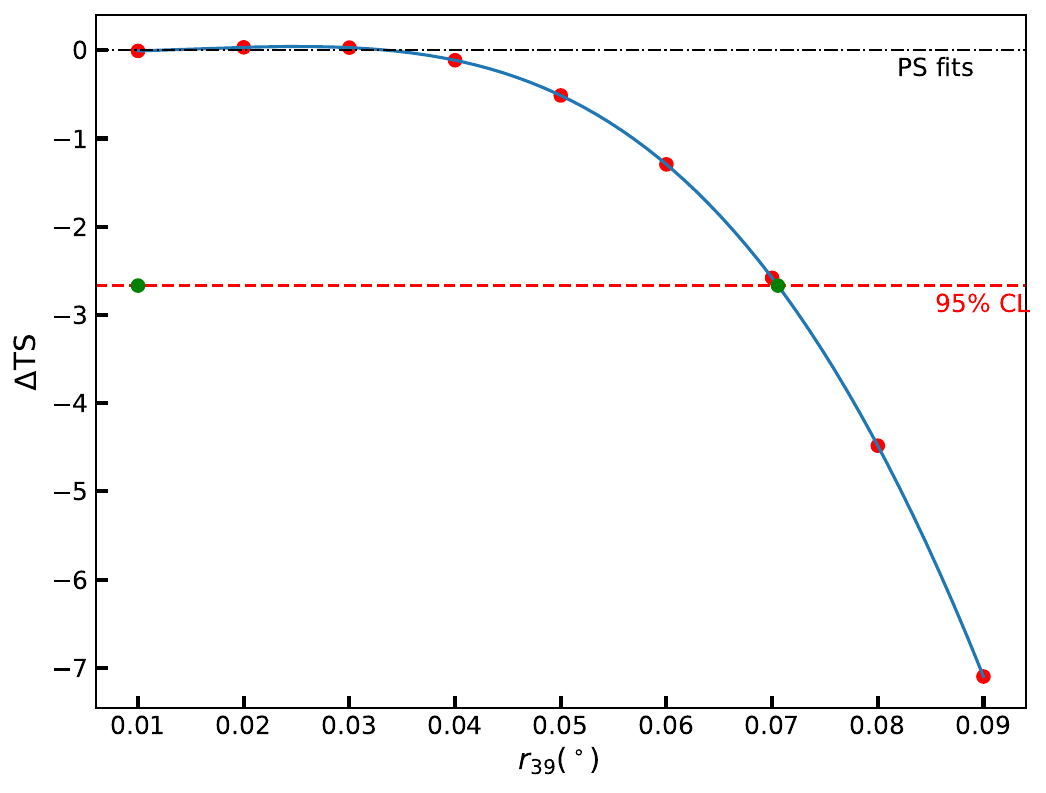}
\caption{The likelihood profile method is used to calculate the upper limit of the source size. Left panel: WCDA ($2-40$ TeV). Right panel: KM2A ($>$ 25 TeV). The red dots represent the TS values obtained for different sizes (after subtracting the TS value of the point-like source model), and the blue solid line is the interpolated smooth curve. The green dot corresponds to the $95\%$ upper limit.} 
\label{fig:Size_uplim}
\end{figure}

\begin{figure}[H]
\centering
\includegraphics[scale=0.32]{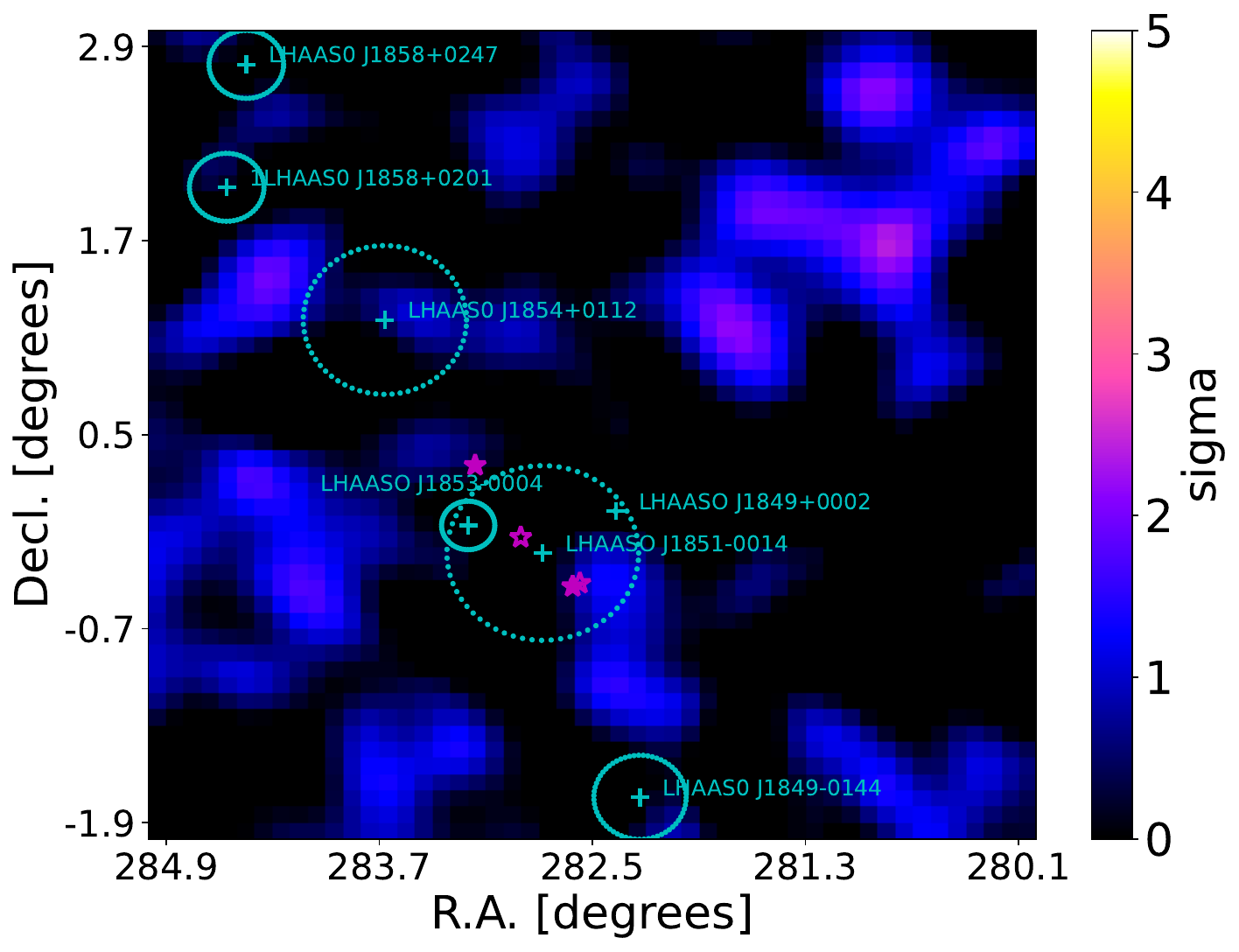}
\includegraphics[scale=0.32]{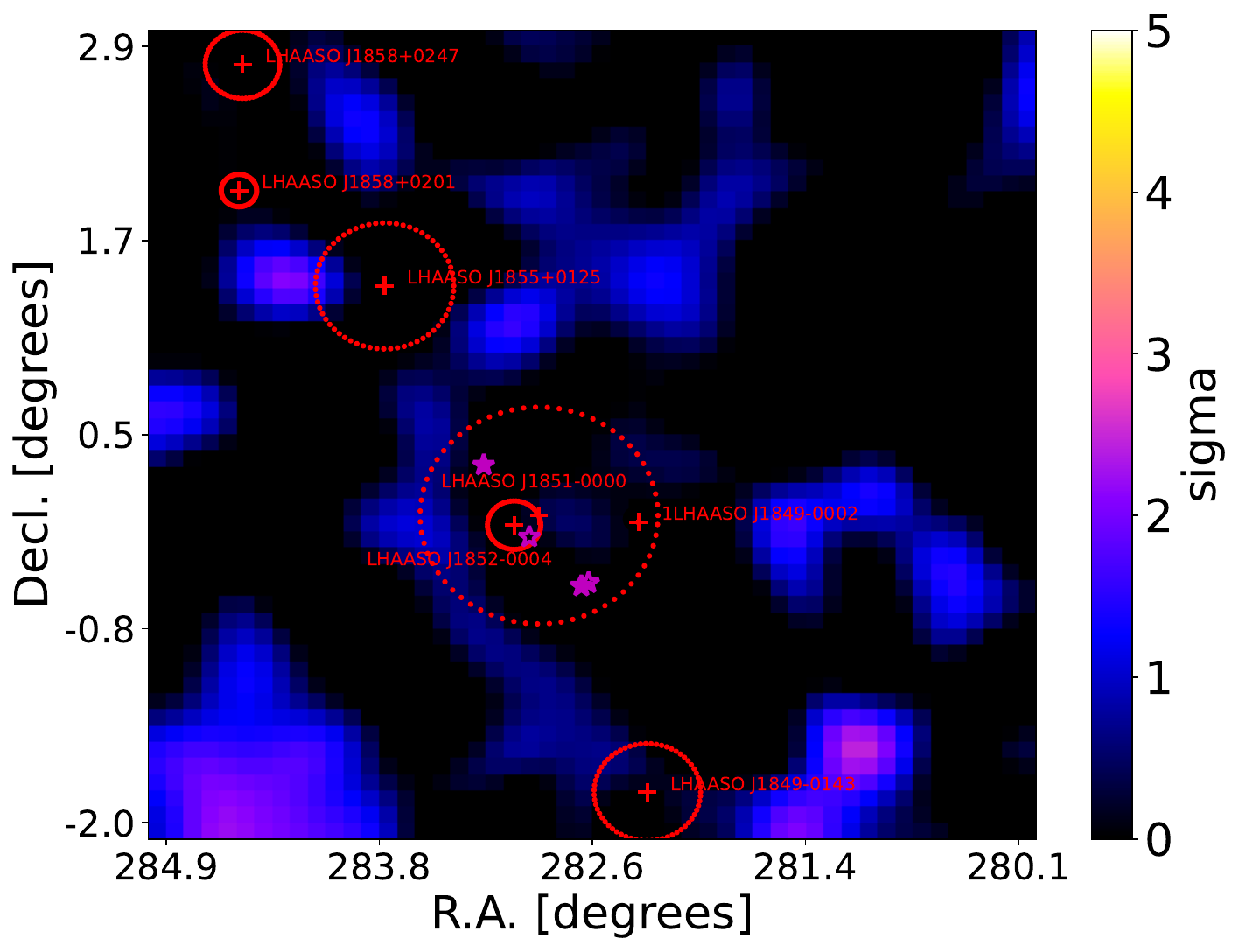}
\caption{Left panel: The residual significance map of WCDA ($2-40$ TeV) within the ROI. Right Panel: The residual significance map of KM2A ($>$ 25 TeV) within the ROI.} 
\label{fig:Remap}
\end{figure}

\begin{figure}[H]
\centering
\includegraphics[scale=0.3]{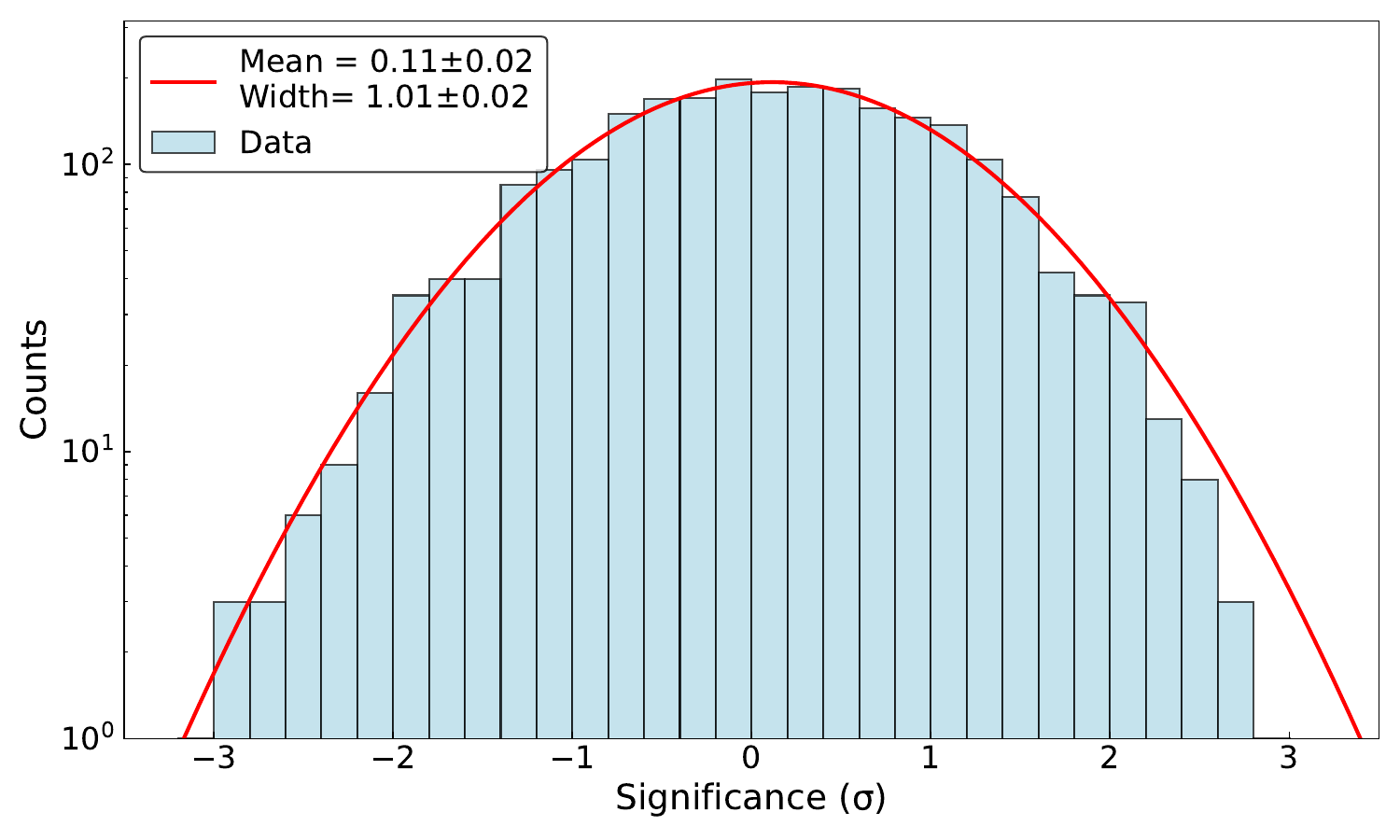}
\includegraphics[scale=0.3]{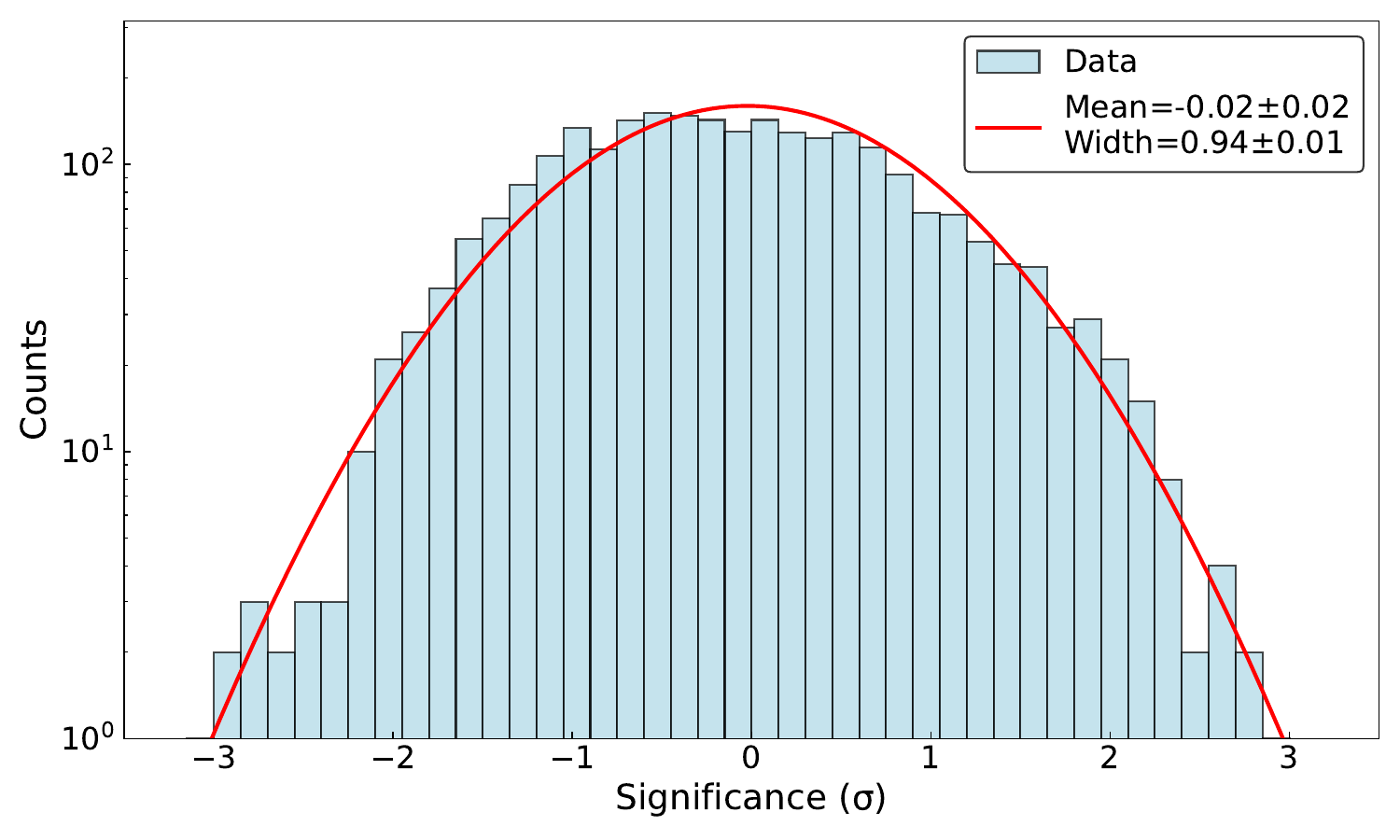}
\caption{The distribution of residual significance map within the ROI for analysis of PSR J1849-0001 at WCDA ($2-40$ TeV) and KM2A ($>$ 25 TeV). In each panel, the red line is the fitting to the distribution with a Gaussian function.} 
\label{fig:1D_Remap}
\end{figure}

\section{The spectrum test of LHAASO J1849-0002}
The spectra of LHAASO J1849-0002 measured by WCDA and KM2A, if described with a power-law function separately, can be given by, 
\begin{equation}\label{eq:spec_src}
    \frac{dN}{dE} = 
    \begin{cases} 
        (5.45 \pm 1.04) \times 10^{-14} \left( \frac{E}{3 \, \text{TeV}} \right)^{-2.07 \pm 0.08} \, \text{TeV}^{-1} \text{cm}^{-2} \text{s}^{-1}, & 2 \, \text{TeV} < E < 40 \, \text{TeV(WCDA)}  \\ 
        (1.53 \pm 0.10) \times 10^{-16} \left( \frac{E}{50 \, \text{TeV}} \right)^{-2.75 \pm 0.06} \, \text{TeV}^{-1} \text{cm}^{-2} \text{s}^{-1}, & > 25 \, \text{TeV(KM2A)} . 
    \end{cases}
\end{equation}
For the all TeV data points, the spectrum can be described by a broken power-law function, see Equation 1 and Figure 2 of the main text. We also performed a joint analysis using WCDA and KM2A data across the entire energy band to evaluate the spatial and spectral models for all sources within the ROI. With the exception of LHAASO J1849-0002, all background sources were modeled with a Gaussian spatial template and a log-parabola (LP) spectrum, with parameters left free to vary. For LHAASO J1849-0002, the spatial model was fixed as a point source based on the separate analysis results. We compared three spectral models for the target: LP, broken power law (BPL), and power law with an exponential cutoff (CPL). The results are summarized in Supplementary Table \ref{tab:jointfit}.
The LP model yielded the best fit.  Consequently, we derived the spectral points using the LP model and compared them with the results obtained from the individual analyses of WCDA and KM2A data, as shown in Supplementary Figure \ref{fig:jointsed}. The results show no significant deviation. We choose to report the results with analyses of WCDA and KM2A data separately. 

\begin{table}[htbp]
\centering
\caption{Joint-fit model parameters of LHAASO J1849-0002}
\begin{threeparttable}
\begin{tabularx}{1\textwidth}{ccccccccc} 
\hline\hline 
Method & \makecell{Spectrum} & $\alpha$ & $\beta$ & $E_{\rm b}$/$E_{\rm cut}$ & $\Delta TS$\tnote{b} & RA & Decl. \\ 
\hline
\multirow{2}{*}{Separate} & \multirow{2}{*}{PL}\tnote{a} & $2.07\pm0.08$ & \multirow{2}{*}{-} & \multirow{2}{*}{-} & \multirow{2}{*}{-} & $282.32\pm0.02$ &$0.03\pm0.02$ \\
& & $2.75\pm0.06$ &  &  &  & $282.24\pm0.01$ &$-0.04\pm0.01$   \\
\hline
\multirow{3}{*}{Joint} & LP & $2.01\pm0.10$ & $0.52\pm0.09$ & - & 0 & $282.27\pm0.01$ &$-0.02\pm0.01$   \\ 
& BPL & $1.94\pm0.14$ & $3.04\pm0.17$ & $62.16\pm18.19$ & -5.9 & $282.27\pm0.01$ &$-0.02\pm0.01$   \\ 
& CPL & $1.96\pm0.09$ & - & $175.91\pm36.36$ & -10.3 & $282.27\pm0.01$ & $-0.02\pm0.01$
\\ 
\hline\hline 
\end{tabularx}
\begin{tablenotes}
\footnotesize
{\item[a] The individual fit results for WCDA (first row) and KM2A (second row), corresponding to Equation \ref{eq:spec_src}.
\item[b] The LP model is used as the baseline, corresponding to $\Delta TS = 0$}
\end{tablenotes}
\end{threeparttable}
\label{tab:jointfit}
\end{table}

\begin{figure}[H]
\centering
\includegraphics[scale=0.3]{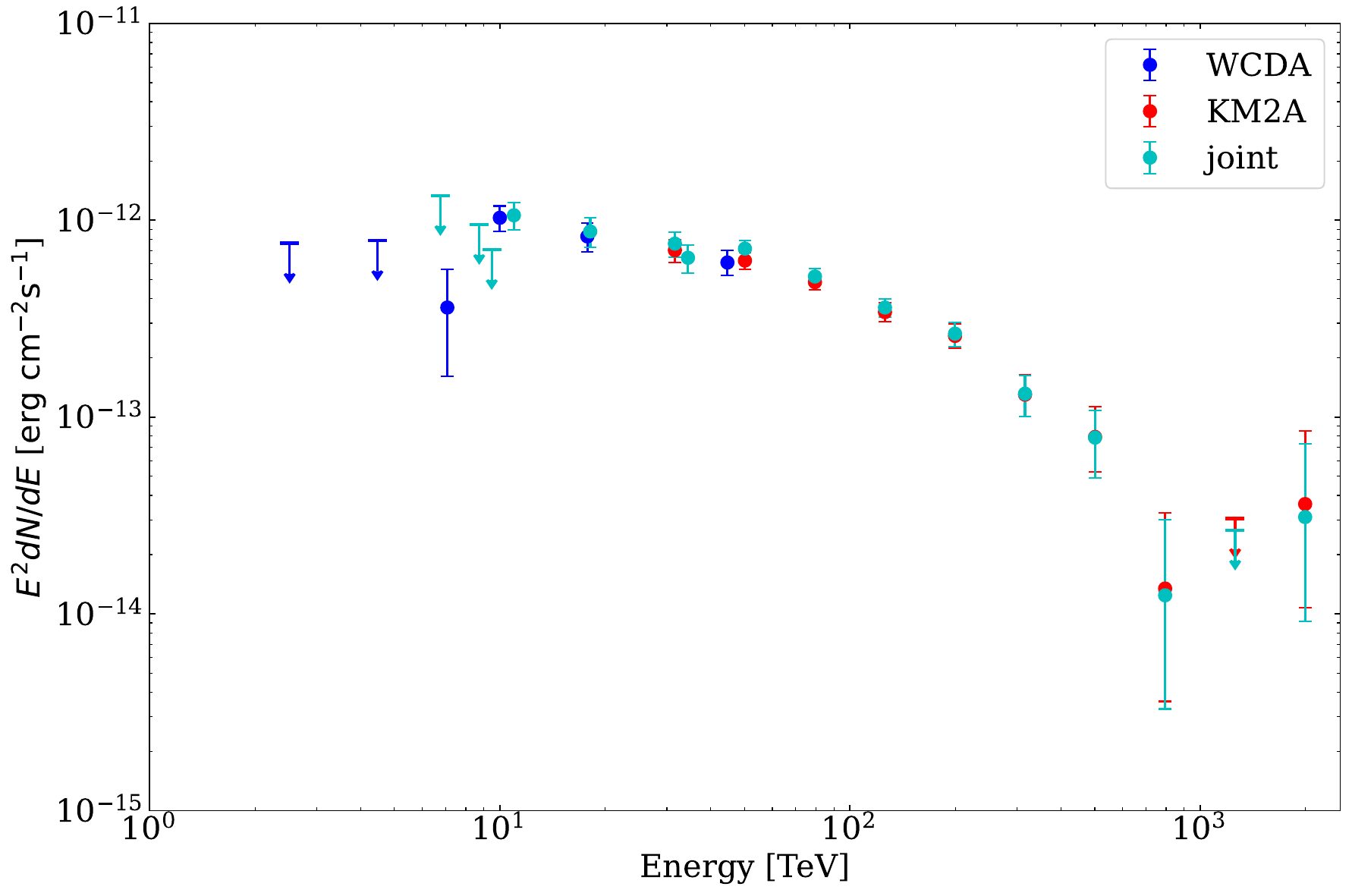}
\caption{Comparison of the spectra obtained from the joint fit and the individual fits of WCDA and KM2A. The joint spectrum is fitted with a log-parabola model, while both individual spectra are fitted with a power-law model.} 
\label{fig:jointsed}
\end{figure}

\section{Probability Estimation for the 2 PeV Photon}
The intensity of GDE is obtained simultaneously in the fitting process, We described its spectrum by a power-law function in the analysis of WCDA and KM2A separately, with the spectral index and normalization factor being set as free parameters. The best-fit average intensity within the ROI is obtained as:
\begin{equation}
    \frac{dN_{\rm GDE}}{dEd\Omega} = 
    \begin{cases} 
        (1.86 \pm 0.12) \times 10^{-10} \left( \frac{E}{3 \, \text{TeV}} \right)^{-2.61 \pm 0.03} \, \text{TeV}^{-1} \text{cm}^{-2} \text{s}^{-1}\text{sr}^{-1}, & 2<~ E<40~\rm TeV\,(WCDA), \\ 
        (5.56 \pm 0.76) \times 10^{-14} \left( \frac{E}{50 \, \text{TeV}} \right)^{-3.47 \pm 0.20} \, \text{TeV}^{-1} \text{cm}^{-2} \text{s}^{-1}\text{sr}^{-1}, &  E>~25~\rm TeV\,(KM2A). 
    \end{cases}
\end{equation}

Supplementary Figure \ref{fig:GDE_SED} show the GDE spectrum obtained in the ROI of LHAASO J1849-0002 and compare it with the GDE spectrum in the inner Galactic plane region ($15^{\circ} < l < 125^{\circ}, |b| < 5^{\circ}$) obtained in our previous work\cite{LHAASO2023_diffuse_sm,LHAASO2025_diffuse_sm}. The results indicate that the GDE intensity in the ROI is significantly higher than the average GDE intensity over the inner Galactic plane region. Note that GDE intensity is supposedly variable from place to place in the Galactic plane, depending on the average CR density and the gas column density in the region. In this study, the ROI center is approximately located at $(l,b)=(35.6^\circ,0.5^\circ)$. This region coincides with the high-intensity region of GDE as can be seen in the 1D longitude profile of GDE reported in Figure 4 of ref \cite{LHAASO2025_diffuse_sm}.

The most energetic photon detected from LHAASO J1849-0002 reaches 2 PeV, with a positional deviation from the source by $0.08^{\circ}$ with an uncertainty of $0.12^{\circ}$. 
We show the shower of this event measured by KM2A in Supplementary Figure~\ref{fig:2PeV_event}.
The chance probability of the event being a cosmic ray (CR) particle is estimated by the number of cosmic ray background. The number of cosmic ray background with energy above 2 PeV is estimated using the Equal-zenith angle method. According to the estimation, there is only 1 photon-like event in 20 background windows with radius of $2^\circ$  (or in a total $251.2^{\circ2}$ background region). If the event in the background window is a true CR event passing our muon cut criteria, the expected CR background in the uncertainty region of the 2\,PeV event is $\pi (0.12)^2/251.2\simeq 0.02\%$. Alternatively, we may estimate the chance probability based on the muon ratio ($N_\mu/N_{\rm e}=10^{-2.47}$) of the 2\,PeV event, with $N_\mu$ and $N_{\rm e}$ being the number of muons and electrons/positrons respectively detected by KM2A in the shower induced by the event. On the other hand, the probability distribution of the muon ratio estimated from high Galactic latitude events (mostly are CR events) of energy $\geq 2\,$PeV shows that the probability of a CR event with $N_\mu/N_{\rm e}\leq 10^{-2.47}$ is $3.29\times 10^{-6}$. In our data, there are total 46 events of energy above 2\,PeV from the direction of the 2\,PeV photon (within a radius of $0.12^\circ$ from the best-fit position). The chance probability of the 2\,PeV event being a CR passing the muon criteria is $46\times 3.26\times 10^{-6}\simeq 0.015\%$, which is consistent with the estimation mentioned above.

\begin{figure}[H]
\centering
\includegraphics[scale=0.42]{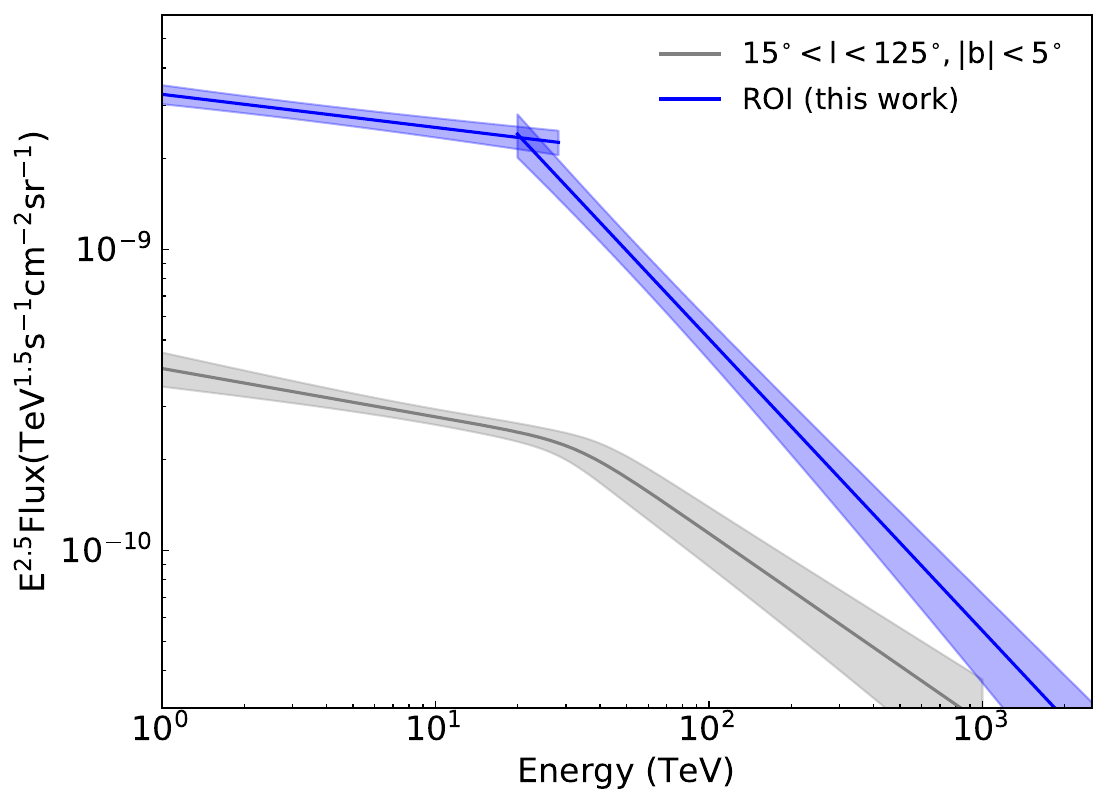}
\caption{GDE spectra measured from ROIs of LHAASO J1849-0002 compared with the GDE in the inner Galactic disk region. The gray error band represents the average GDE flux in the inner Galactic plane region ($15^{\circ} < l < 125^{\circ}$ and $|b| < 5^{\circ}$) given by ref.\cite{LHAASO2025_diffuse_sm}.} 
\label{fig:GDE_SED}
\end{figure}

There is also a chance probability that the 2\,PeV photon is from GDE. 
Based on the best-fit GDE flux within the uncertainty ($0.12^\circ$) of the photon's arrival direction $f_{\rm GDE}(E)$ and the exposure of LHAASO-KM2A for 2\,PeV photon events $W(E;2\,\rm PeV)$, we may estimate the chance probability. More specifically, $f_{\rm GDE}$ is calculated by $dN/dEd\Omega \times \pi (0.12^\circ)^2 \times q(l, b)$, where $q(l,b)$ is the ratio of the GDE intensity at the position of the source to the average GDE intensity within the ROI. $W(E;2\,{\rm PeV})=A_{\rm eff}(l,b)T_{\rm eff}(l,b)k(E;2\,\rm PeV)$, where $A_{\rm eff}(E,l,b)$ and $T_{\rm eff}(l,b)$ are, respectively, the effective area and the effective exposure time of LHAASO-KM2A toward the photon's arrival direction, and $k(E;2\,\rm PeV)$ is the probability of a photon of true energy $E$ being reconstructed as a photon of energy from $10^{3.2}\,$TeV to $10^{3.4}\,$TeV. The chance possibility of detecting a 2\,PeV photon event of GDE origin can then be given by $\int{f_{\rm GDE}(E)W(E;{\rm 2\,PeV}) dE} \simeq 0.03 \%$. 

In addition, a large extended source LHAASO J1851-0000 overlap LHAASO J1849-0002. We evaluated the systematic uncertainty arising from the modeling of this close-by source.  To do this, we fixed the positions and extensions of all other sources (except LHAASO~J1851-0000 and LHAASO~J1849-0002) in ROI to their original best-fit values. Then we fixed the position of LHAASO~J1851-0000 to its best-fit value but varying it spatial extension to the $90\%$ C.L. upper and lower bounds. The target source (LHAASO~J1849-0002) was then re-fitted under these conditions. We found that the resulting deviation in the target source's position was less than $0.012^\circ$ ($0.015^\circ$) for WCDA (KM2A) with respect to its original best-fit position, and the systematic uncertainty on the flux, derived from the re-fitted spectra, was less than $8.9\%$ ($7.4\%$) for WCDA (KM2A). The energy spectrum measured by LHAASO J1851-0000 above 25\,TeV is described by a power-law function: $dN/dE = (2.08\pm0.38)\times10^{-16}(E/50~\rm TeV)^{-3.39\pm0.13}~\rm TeV^{-1}cm^{-2}s^{-1}$. Combining this spectrum with the instrument's exposure, we estimated the possible contribution of 2\,PeV photon from  this source.
Regarding the spatial distribution, the source was modeled with a 2D symmetric Gaussian template with $\sigma=0.67^\circ$ (see Supplementary Table~\ref{tab:morfit_km2a}). The position of the 2\,PeV photon is at a radial distance of $r = 0.55^\circ$ away from the source center, and we obtained the spatial probability density, $S(r=0.55^\circ) = 0.25~\text{deg}^{-2}$, based on the Gaussian template. The spectrum per solid angle of the source at the position of the 2\,PeV photon can be given by $dN/dEd\Omega=S(r)dN/dE$. Then, same as the estimation for the GDE contribution shown above the probability of the 2\,PeV photon being from LHAASO~J1851-0000 can be given by $\pi (0.12^\circ)^2\times \int (dN/dEd\Omega)W(E;2{\rm PeV})dE\simeq 0.1\%$. 

For a more conservative estimate of $\eta$, we note that even after excluding the highest-energy photon, the gamma-ray emission still extends up to 700 TeV. Based on energy conservation, the parent electrons must possess an energy of at least 700 TeV, which implies an acceleration efficiency of $\eta > 0.09$. If we follow the energy relationship between the up-scattered photon and the parent electron with CMB as the target radiation field\cite{LHAASO2021_Crab_sm}, $E_{\rm e} \approx 2.15(E_{\gamma}/1~\rm PeV)^{0.77}~\rm PeV$, electrons of energy $E_{\rm e} \approx 1.6~\rm PeV$ are needed, resulting in $\eta > 0.21$.

\begin{figure}[H]
\centering
\includegraphics[scale=0.8]{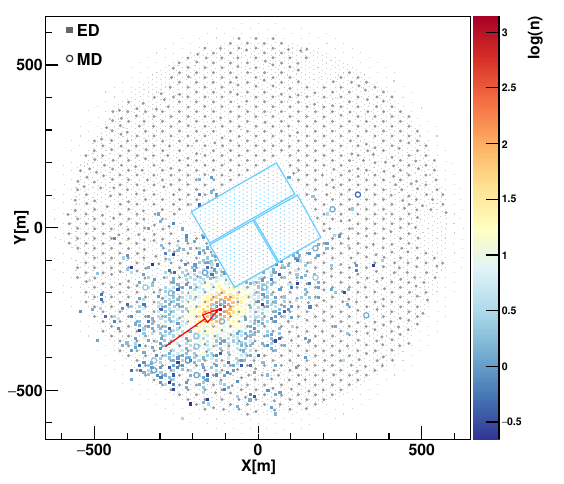}
\caption{The 2\,PeV photon event detected by LHAASO-KM2A. About 8650 electron/positron events are detected in the shower, whereas only 29 muon events are detected, suggesting this event a photon. The positional deviation of the 2\,PeV photon from the source by $0.08^\circ\pm 0.12^\circ$.} 
\label{fig:2PeV_event}
\end{figure}

\section{Fermi-LAT data}
This section presents the TS map for LHAASO J1849-0002 derived from Fermi-LAT data. The analysis configuration, including the event selection and background model, is identical to that described in method of the main manuscript.

\begin{figure}[H]
\centering
\includegraphics[scale=0.38]{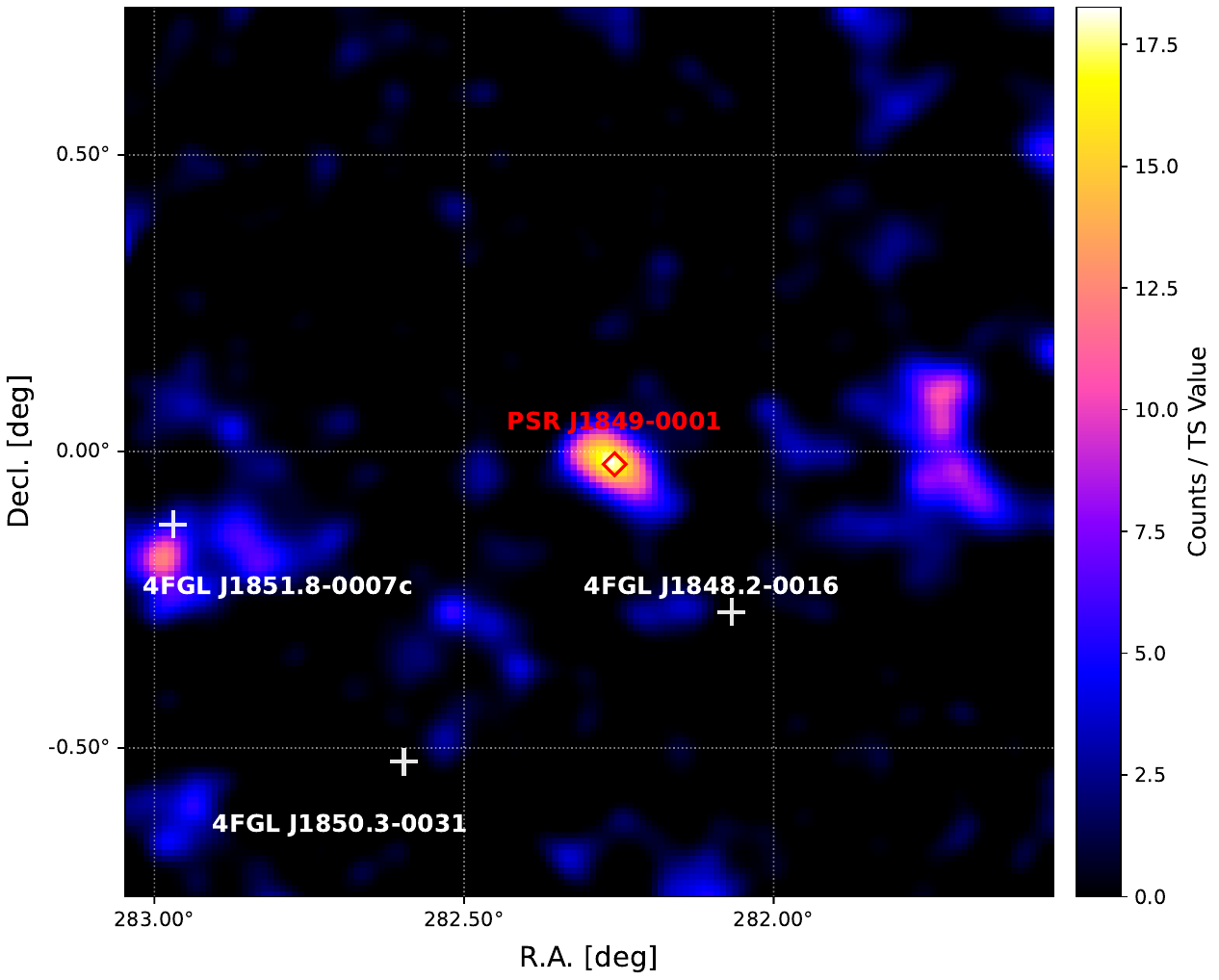}
\caption{The $>$10 GeV residual map of the Fermi-LAT. 
The red prismatic is the positions of PSR J1849-0001. 
The white plus marks the location of the 4FGL sources.
} 
\label{fig:fermi}
\end{figure}

\section{Systematic tests on the impact of X-ray background spatial variations}

The Galactic background varies with Galactic longitude and latitude. Since the background region is relatively far from the source region in our analysis, we tested the possible influence of the spatial variation of background on the result.  We first separated the background region into two parts according to their Galactic longitude, and used each part in spectral fitting separately to examine how they affects our results. By doing so, we found the source photon index with higher longitude background region to be \( \Gamma = 2.07\pm0.26 \), and the source photon index obtained with lower longitude background to be \( \Gamma = 2.43\pm0.24 \). Similarly, we also divided the background region into higher latitude part and lower latitude part, resulting in \( \Gamma = 2.29\pm0.26 \) and \( \Gamma = 2.47\pm0.27 \) respectively. All these source results consistent with \( \Gamma = 2.31\pm0.23 \), which is obtained from the simultaneously spectral fitting with the entire background region. This suggests that spatial inhomogeneity in the Galactic background will not introduce a significant impact on our analysis.

\section{One-zone leptonic model}
During propagation, a fraction of photons with energies exceeding 100 TeV are absorbed by the background photon fields (ISRF and CMB). We account for this attenuation effect in our analysis. Supplementary Figure \ref{fig:SED_absorb} shows the resulting spectrum after applying the absorption correction based on the optical depth of the region.

\begin{figure}[H]
\centering
\includegraphics[width=0.5\textwidth]{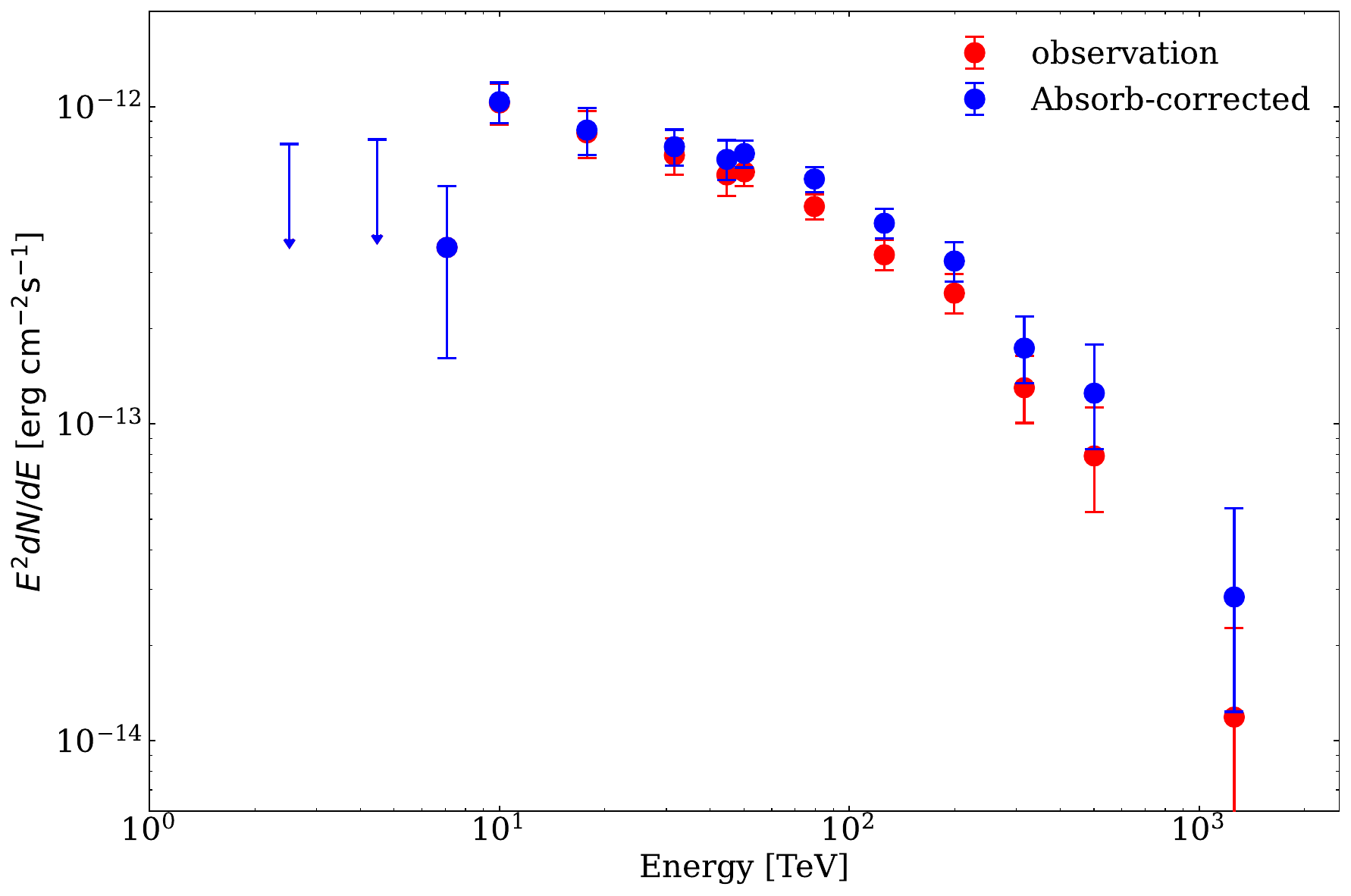}
\caption{Spectrum of LHAASO~J1849-0002 corrected for the absorption by ISRF and CMB with a source distance of 7\,kpc, shown as blue points. Red points represent the observed spectrum. The last three energy bins are combined. Error bars represent $1\sigma$ uncertainties, and arrows indicate $95\%$ upper limits.
} 
\label{fig:SED_absorb}
\end{figure}

We model the $\gamma$-ray SED of the 
LHAASO J1849-0002 with an inverse Compton scattering (ICS) model of $e^{\pm}$ diffusing out from the pulsar, attempted to model the multiwavelength spectrum with a simple one-zone (leptonic) radiation model using the Python package naima \cite{2015ICRC...34..922Z_sm}. The electron spectrum is modeled using a BPL function,
\begin{equation}
\frac{dN_{\rm e}}{dE_{\rm e}} = A \times
\begin{cases} 
 \left( \frac{E_{\rm e}}{10~\rm GeV} \right)^{-s_1} & \text{if~} E \leq E_{\text{e,br}}, \\ 
 \left( \frac{E_{\text{br}}}{10~\rm GeV} \right)^{s_2 - s_1} \left( \frac{E_{\rm e}}{10~\rm GeV} \right)^{-s_2} & \text{if~} E > E_{\text{e,br}}. 
\end{cases}
\end{equation}
The normalisation $A$, spectral index $s_1$, $s_2$, the break energy $E_{\rm e,br}$, and the ambient magnetic field B are treated as free parameters in the MCMC fitting. The best-fit values are $A = 6.32^{+2.26}_{-2.33}\times10^{38}~\rm eV^{-1}$, $s_1 = 2.34^{+0.04}_{-0.06}$, $s_2 = 3.29 ^{+0.41}_{-0.26}$ and $E_{\rm e,br} = 209.58 ^{+69.92}_{-60.87} \rm TeV$ and $B = 2.81\pm0.24 \rm \, \mu G$. In Supplementary Figure \ref{fig:SED_ele}, we present the best-fit electron distribution. The multi-wavelength spectral fitting results are presented in the left panel of Supplementary Figure \ref{fig:SED_fit}, where the dashed lines represent the Inverse Compton emission components arising from different seed photon fields. The right panel of Supplementary Figure \ref{fig:SED_fit} displays the corner plot of the best-fit parameters. For comparison, we also performed the fitting using a CPL function for the electron energy distribution with $A = 4.94^{+7.21}_{-2.09}\times10^{38}~\rm eV^{-1}$, $\alpha=2.30^{+0.10}_{-0.07}$, $E_{\rm cut}= 640.05^{+271.14}_{-173.16}~\rm TeV$ and $B = 2.85\pm 0.25~\mu G$; these results are shown in Supplementary Figure \ref{fig:SED_fit_CPL}. The fits yield $\chi^2$ values of 10.10 and 11.86 for the BPL and CPL models, corresponding to reduced chi-squared values ($\chi^2_\nu$) of 1.01 and 1.08, respectively. The two models give equally good fitting. Note that, using the CPL model does not change our conclusion, because whether or not a break presents around 200\,TeV or a cutoff appears around $650\,$TeV in the electron spectrum, we need acceleration of $>2$\,PeV electrons in the PWN to explain the observed 2\,PeV photon. 

\begin{figure}[H]
\centering
\includegraphics[scale=0.3]{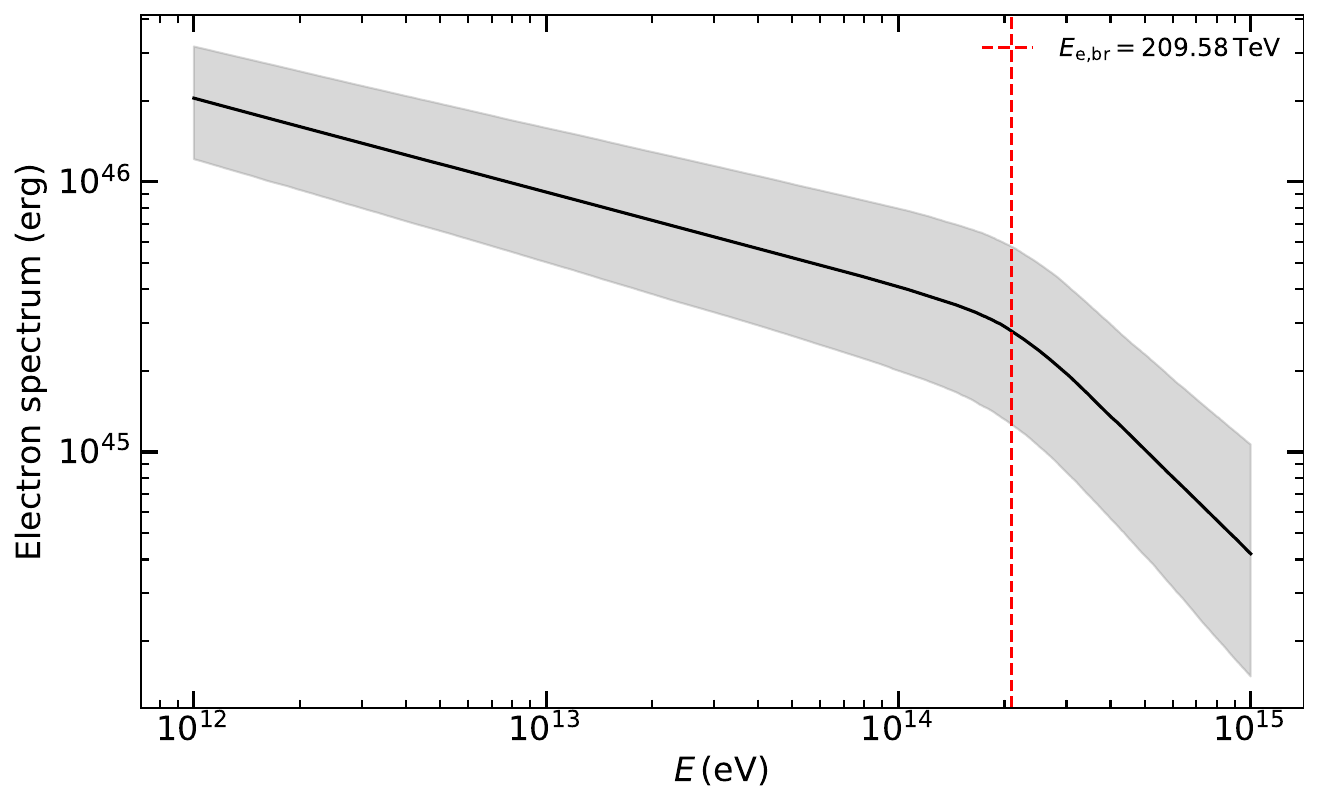}
\caption{The best-fit electron spectrum obtained by NAIMA when fitting multi-wavelength emission through synchrotron radiation and IC processes.} 
\label{fig:SED_ele}
\end{figure}

\begin{figure}[H]
\centering
\includegraphics[scale=0.45]{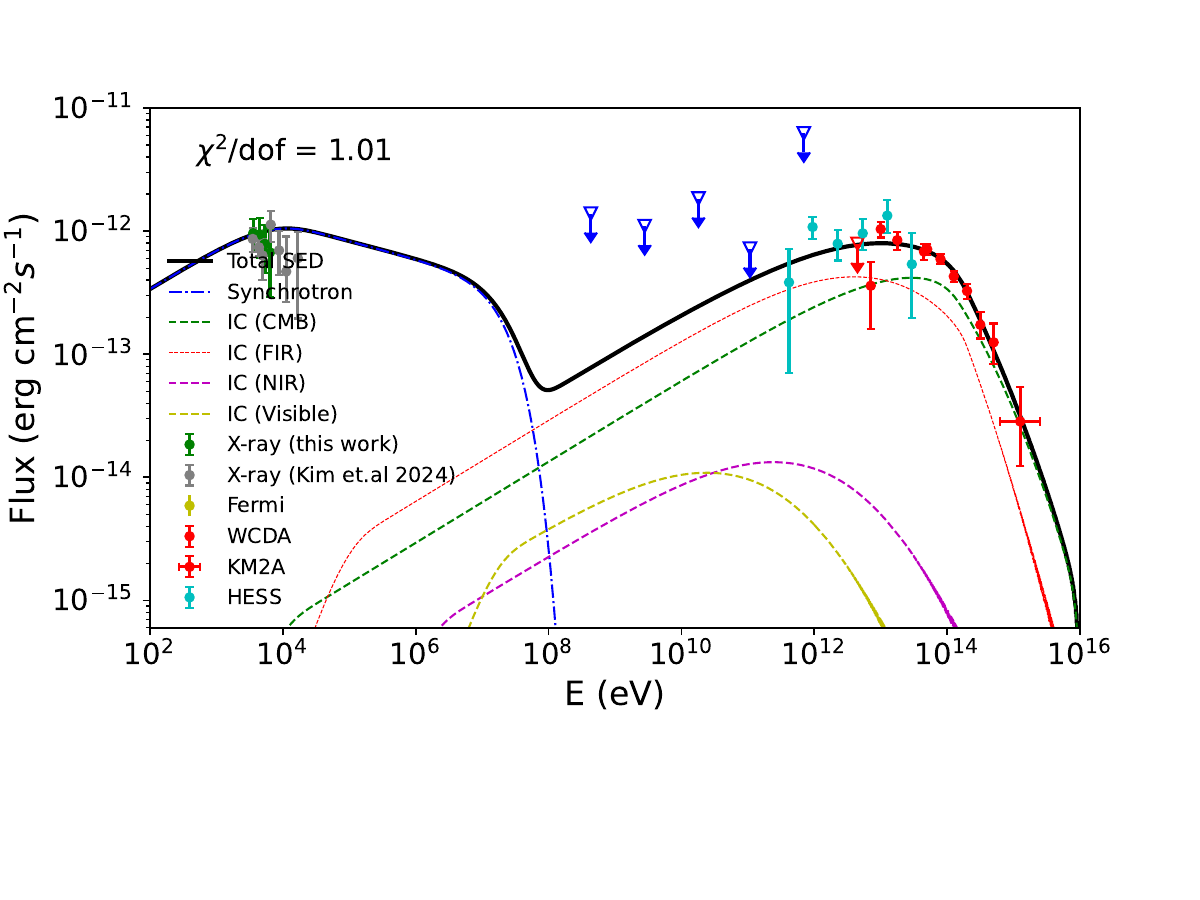}
\includegraphics[scale=0.26]{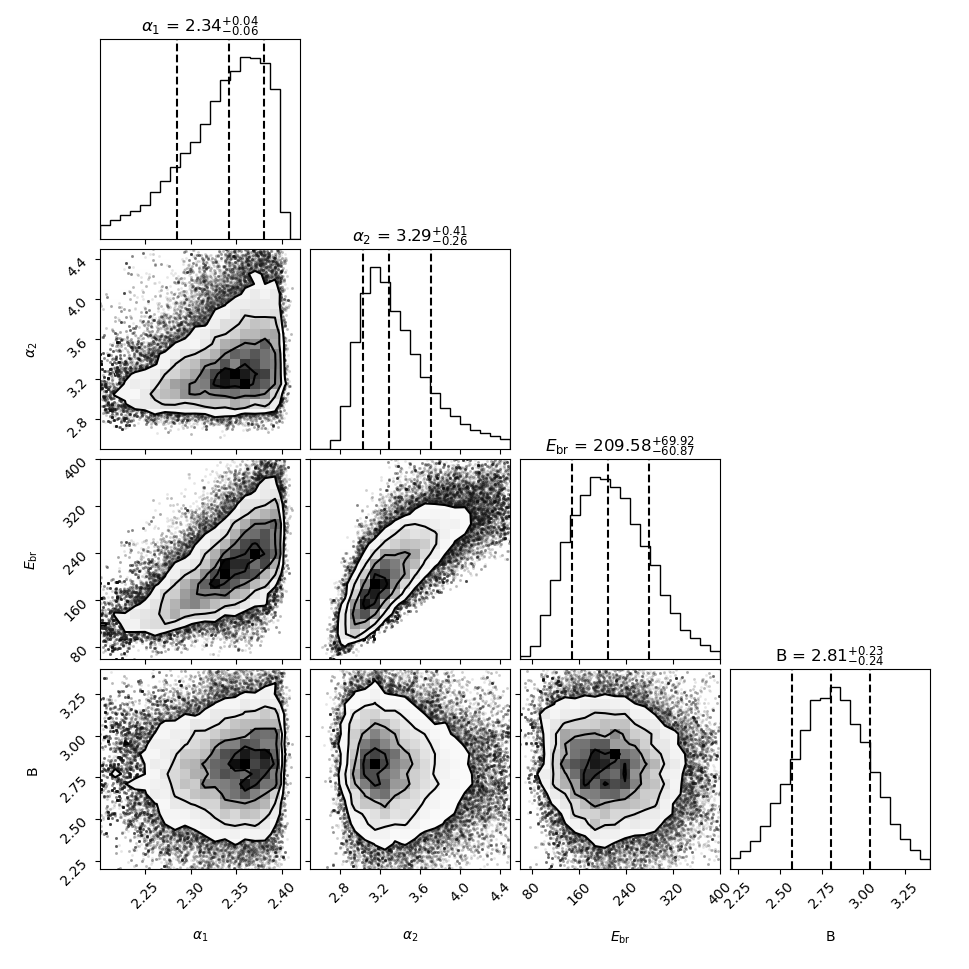}
\caption{The multiwavelenth SED fitting with the BPL spectrum and the corresponding best-fit parameters.} 
\label{fig:SED_fit}
\end{figure}

\begin{figure}[H]
\centering
\includegraphics[scale=0.45]{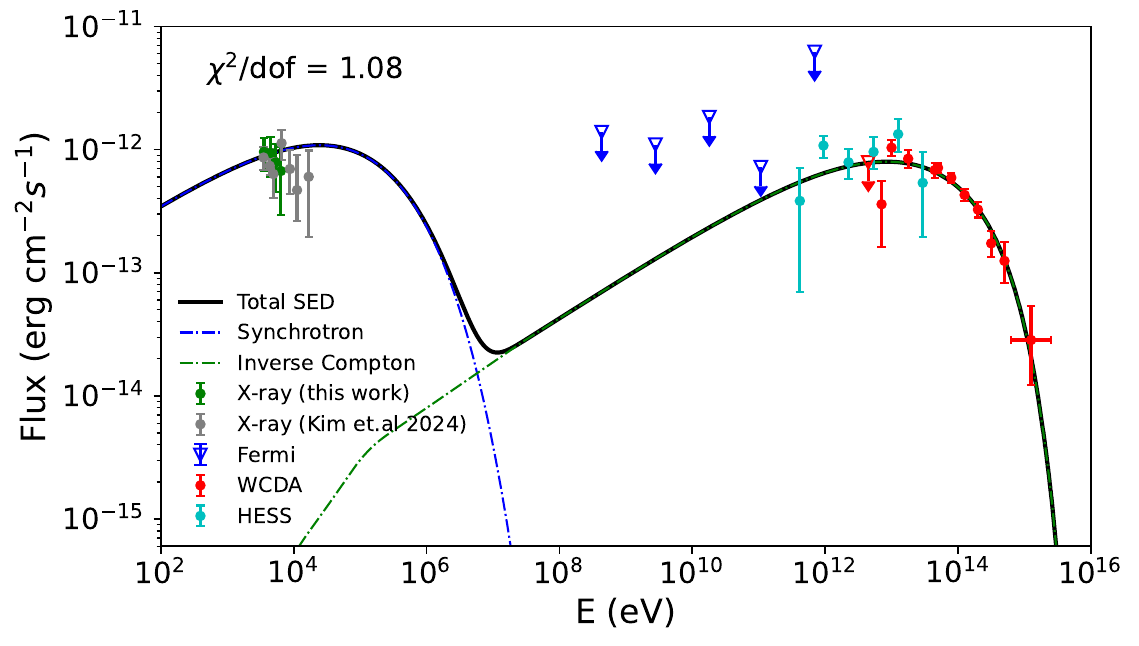}
\includegraphics[scale=0.3]{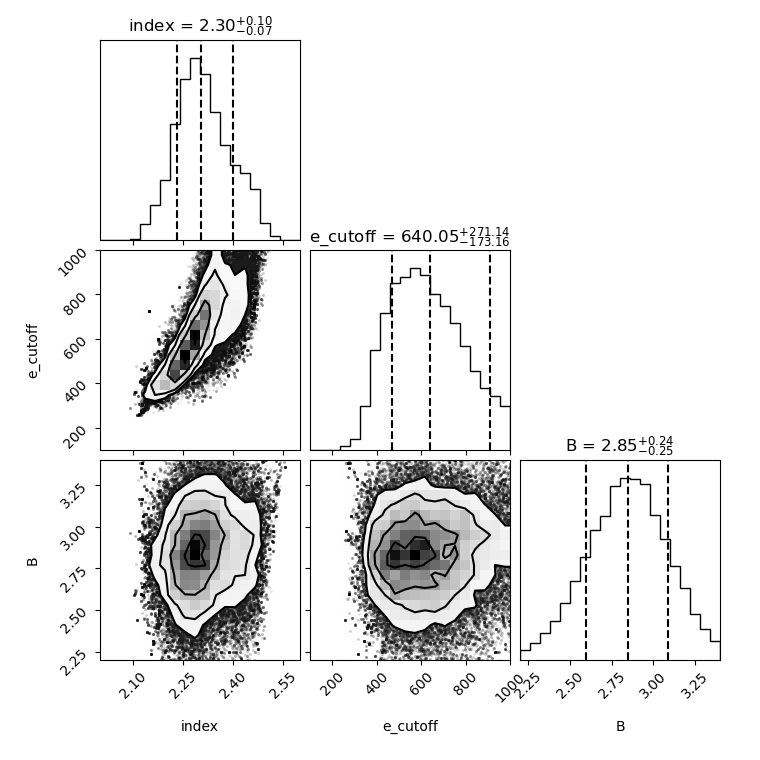}
\caption{The multiwavelenth SED fitting with the CPL spectrum and the corresponding best-fit parameters. }
\label{fig:SED_fit_CPL}
\end{figure}

\section{Influence of the spindown of the pulsar}

\begin{figure}[H]
    \centering
    \includegraphics[width=0.6\textwidth]{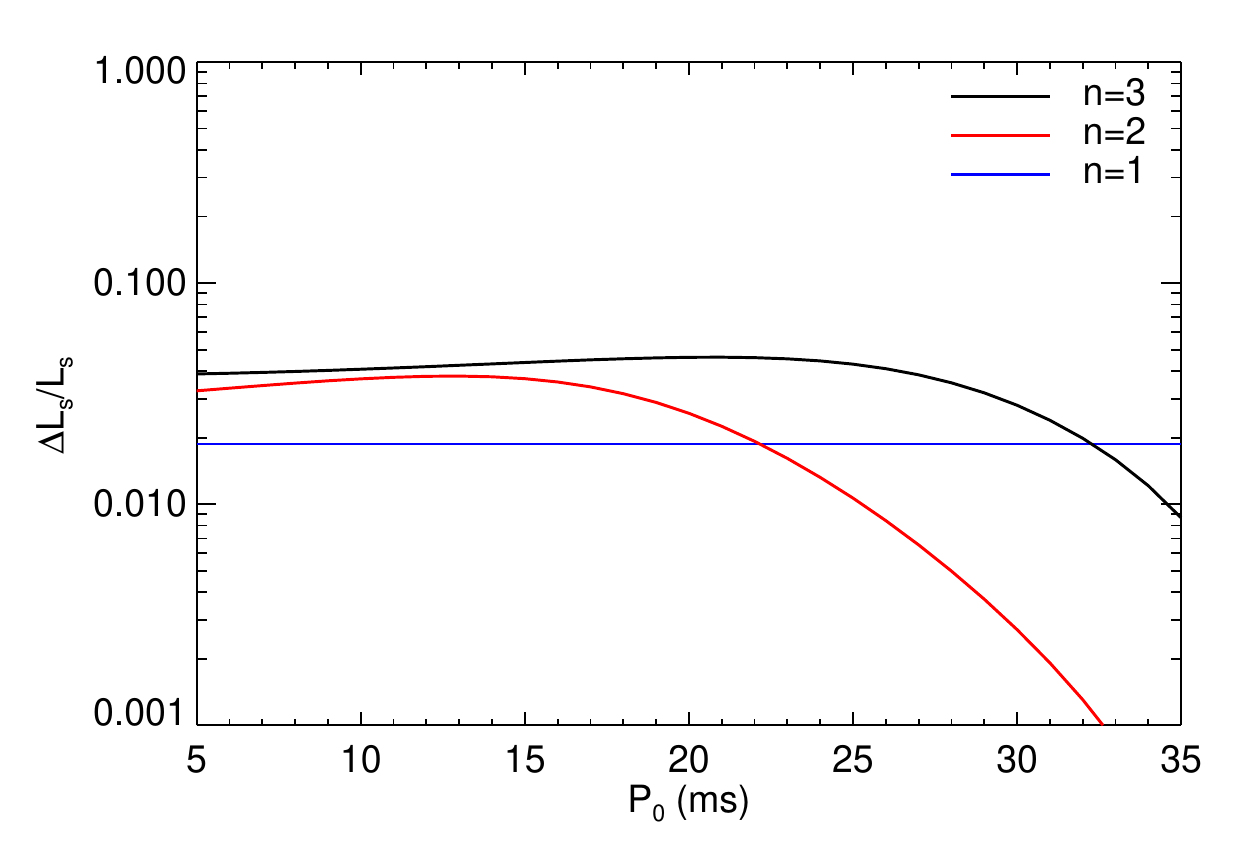}
    \caption{Ratio of increased spindown luminosity at 800 years ago to the present spindown luminosity of PSR~J1849-0001 for different initial rotational period $P_0$ and braking index $n$.}
    \label{fig:spindown}
\end{figure}

\par
\vspace{3em}

\bibliography{ms_sm}

\end{document}